\documentclass[lettersize,journal,twoside]{IEEEtran}
\usepackage{amsthm}
\usepackage{amsmath,amsfonts}
\usepackage{algorithmic}
\usepackage{algorithm}
\usepackage{array}
\usepackage[backref]{hyperref}
\usepackage{booktabs}
\usepackage{bm}
\usepackage[caption=false,font=footnotesize,labelfont=rm,textfont=rm]{subfig}
\usepackage{mathtools}
\usepackage{setspace}
\usepackage{textcomp}
\usepackage{threeparttable}
\usepackage{makecell}
\usepackage{stfloats}
\usepackage{url}
\usepackage{verbatim}
\usepackage{graphicx}
\usepackage{cite}
\newcommand*{\dif}{\mathop{}\!\mathrm{d}}
\newcommand{\T}{\mathrm{T}}

\begin{document}

\title{Dynamic Modeling and Stability Analysis for Repeated LVRT Process of Wind Turbine \\ Based on Switched System Theory}

\author{
	Qiping Lai,~\IEEEmembership{Student Member,~IEEE}, Chen Shen,~\IEEEmembership{Senior Member,~IEEE}, and Dongsheng Li,~\IEEEmembership{Student Member,~IEEE}
	\thanks{This work was supported by the The National Key R\&D Program of China ``Response-driven intelligent enhanced analysis and control for bulk power system stability'' under Grant 2021YFB2400800. \textit{(Corresponding author: Chen Shen.)}}
	\thanks{The authors are with the Department of Electrical Engineering, Tsinghua University, Beijing 100084, China (e-mail: lqp22@mails.tsinghua.edu.cn; shenchen@mail.tsinghua.edu.cn; lids19@mails.tsinghua.edu.cn).}
}



\maketitle
\begin{abstract}
The significant electrical distance between wind power collection points and the main grid poses challenges for weak grid-connected wind power systems. A new type of voltage oscillation phenomenon induced by repeated low voltage ride-through (LVRT) of the wind turbine has been observed, threatening the safe and stable operation of such power systems. Therefore, exploring dynamic evolution mechanisms and developing stability analysis approaches for this phenomenon have become pressing imperatives. This paper introduces switched system theory for dynamic modeling, mechanism elucidation, and stability analysis of the repeated LVRT process. Firstly, considering the external connection impedance and internal control dynamics, a novel wind turbine grid-side converter (WT-GSC) switched system model is established to quantitatively characterize the evolution dynamic and mechanism of the voltage oscillation. Subsequently, a sufficient stability criterion and index grounded in the common Lyapunov function are proposed for stability analysis and assessment of the WT-GSC switched system. Moreover, to enhance the system stability, the Sobol' global sensitivity analysis method is adopted to identify dominant parameters, which can be further optimized via the particle swarm optimization (PSO) algorithm. Finally, simulations conducted on a modified IEEE 39-bus test system verify the effectiveness of the proposed dynamic modeling and stability analysis methods.
\end{abstract}

\begin{IEEEkeywords}
Grid-connected wind power system, repeated LVRT-caused voltage oscillation, switched system, dynamic evolution mechanism, stability criterion, stability index, sensitivity analysis, parameter optimization.
\end{IEEEkeywords}

\section{Introduction}

\setstretch{0.91}

\IEEEPARstart{N}{owadays}, large-scale wind power is transmitted over long distances to load centers via high-voltage AC/DC power transmission systems \cite{Ref1_global_wind_report_2023, Ref2_WEGS_overview_2016}. The wind power collection sending end system is electrically far away from the main grid, lacking the support of synchronous generators, so that shows distinct weak grid characteristics \cite{Ref3_weak_grid_reactive_voltage_2013}. A weak power grid, with a short circuit ratio (SCR) generally less than 2.0, is susceptible to large voltage fluctuations and prone to collapse during faults \cite{Ref4_short_circuit_ratio_SunHuadong_2022}. It is vulnerable to voltage oscillations, transient overvoltages, and other stability problems, affecting power systems' safety and stability \cite{Ref5_weak_grid_transient_voltage_SunDawei_2021, Ref6_weak_grid_SSI_2017}. Recently, it has been observed in weak grid-connected wind power systems that the grid-connected point voltage oscillates with the wind turbine repeatedly entering and exiting low voltage ride-through (LVRT), even leading to large-scale wind turbines off-grid. This new phenomenon differs significantly in forms and mechanisms from stochastic voltage oscillations stemming from the randomness and fluctuation of wind power \cite{Ref7_Jibei_voltage_fluctuation_XuMan_2021, Ref8_LVRT_voltage_fluctuation_WuLinlin_2022}. Consequently, it is of vital importance to study the dynamic evolution mechanisms and stability analysis approaches for this voltage oscillation phenomenon.

There have been some studies about the grid-connected point voltage oscillation phenomenon of wind turbines connected to weak grids. Reference \cite{Ref10_abnormal_transient_performance_LanTiankai_2023} revealed two typical LVRT-dominated transient performances of distributed renewable energies, i.e., repeated LVRT and continuous LVRT, and repeated LVRT may cause voltage oscillations. Reference \cite{Ref9_dynamic_voltage_monotone_control_GanDeqiang_2021} explained the repeated LVRT problem within the monotone control theory framework, identifying the low post-fault voltage level as the direct cause of this problem. In \cite{Ref7_Jibei_voltage_fluctuation_XuMan_2021} and \cite{Ref8_LVRT_voltage_fluctuation_WuLinlin_2022}, the mechanism of repeated LVRT-caused voltage oscillations was elucidated. When the voltage drops below the LVRT threshold, the wind turbine will work under the LVRT condition, prioritizing reactive power compensation while reducing the active power output. The voltage will rise through the reactive power priority control, until the wind turbine returns to normal operation. If this cycle keeps repeating, the grid-connected point voltage oscillation will occur. Reference \cite{Ref11_LVRT_voltage_influence_QiJinling_2023} indicated that reactive power control switching and reactive power recovery strategies are the roots of repeated voltage fluctuations. Furthermore, an optimal control strategy of LVRT was also proposed to achieve fast and stable voltage recovery. As analyzed in \cite{Ref12_bifurcation_analysis_XueAncheng_2023}, the voltage oscillation phenomenon corresponds to high output excited low voltage ride-through control induced non-smooth bifurcation (HLINB), and the critical criterion of HLINB was given based on the second-order cone algorithm.

The aforementioned research \cite{Ref7_Jibei_voltage_fluctuation_XuMan_2021, Ref8_LVRT_voltage_fluctuation_WuLinlin_2022, Ref9_dynamic_voltage_monotone_control_GanDeqiang_2021, Ref10_abnormal_transient_performance_LanTiankai_2023, Ref11_LVRT_voltage_influence_QiJinling_2023, Ref12_bifurcation_analysis_XueAncheng_2023} is still based on steady-state power flow calculation and static voltage stability limit analysis, employing the active power-voltage (P-V) curve to illustrate the voltage oscillation phenomenon. However, power flow equations are algebraic and only reflect the distribution of source and load power injections in the grid \cite{Ref13_power_system_analysis_Hadi_2011, Ref14_power_system_analysis_Glover_2022}, thus attributing voltage oscillations to repeated wind power fluctuations. Therefore, the above mechanisms cannot accurately capture the dynamic evolution of grid-connected point voltage, current, and power when the wind turbine repeatedly enters and exits LVRT. It is necessary to consider the internal differential control dynamics of the wind turbine, which can account for its dynamic behavior and stability during the repeated LVRT process.

The reactive power control strategies will switch when the wind turbine enters and exits LVRT \cite{Ref15_GBT_wind_farm_specification_2021}. This switching control system contains continuous dynamic processes and discrete switching events \cite{Ref16_WPGS_modelling_PhD_Fan_2010}. Thus, switched system theory can be introduced to characterize the dynamic voltage evolution and analyze the stability of grid-connected wind power systems. Hybrid systems are dynamic systems with hierarchical structures whose states and outputs depend on continuous dynamics, discrete events, and their interactions \cite{Ref36_challenges_to_control_1987, Ref37_C2_theory_1987, Ref38_HS_power_system_IA_2000}. The switched system, a significant category of hybrid systems, consists of a family of continuous-time subsystems and discrete switching laws indicating the active subsystem at each instant of time \cite{Ref17_switched_system_PhD_Xu_2001, Ref18_network_control_system_2012}. According to different switching modes, switched systems can be classified as state-dependent or time-dependent \cite{Ref18_network_control_system_2012, Ref19_switching_in_systems_Liberzon_2003}. The modeling procedures and system state trajectory-based dynamic evolution analysis methods for switched systems are introduced in \cite{Ref20_trajectory_analysis_IA_2000, Ref21_DCDC_switched_linear_systems_HuZongbo_2005, Ref22_guishengzhaiAnalysisDesignSwitched2006}. The last decades have seen extensive research on stability analysis of switched systems, which can be roughly divided into two main problems: one is to find the stability criteria for switched systems under given or arbitrary switching laws; the other is to identify switching laws making switched systems asymptotically stable \cite{Ref19_switching_in_systems_Liberzon_2003 ,Ref23_D.LiberzonBasicProblemsStability1999, Ref24_zhendongsunSwitchedLinearSystems2005}. Powerful analysis tools such as common Lyapunov functions and multiple Lyapunov functions have been developed to address these problems \cite{Ref19_switching_in_systems_Liberzon_2003, Ref25_m.s.branickyMultipleLyapunovFunctions1998, Ref26_robertshortenStabilityCriteriaSwitched2007}. Other active research fields, including observability \cite{Ref27_AneelTanwaniObservabilitySwitchedLinear2013}, controllability \cite{Ref28_GuilhermeS.VicinansaControllabilityFiniteDataRate2023}, reachability \cite{Ref29_ZhendongSunControllabilityReachabilityCriteria2002}, feedback stabilization \cite{Ref30_RaphaelM.JungersFeedbackStabilizationLinear2017}, and optimal control \cite{Ref31_GorgesOptimalControlScheduling2011} for switched systems, are also in the ascendant.

In summary, existing studies have pointed out that the switching of power control strategies is the root cause of voltage oscillations induced by repeated LVRT. However, there is still a lack of quantitative analysis relating this phenomenon to the dynamic control parameters of the wind turbine. Therefore, this paper introduces switched system theory for model construction, mechanism elucidation, and stability analysis of the LVRT switching process, providing a dynamic and quantitative perspective on these issues. The main contributions are listed as follows:
\begin{enumerate}
	\item {Considering the wind turbine's external connection impedance and internal control dynamics, a novel switched system model of the wind turbine grid-side converter (WT-GSC) is constructed on the basis of its structure and control strategies. This model characterizes the switching dynamics of the repeated LVRT process.}
	\item {By substituting actual parameters into the established WT-GSC switched system model, the dynamic variations of the grid-connected point voltage, current, and power during the repeated LVRT process are captured, illustrating the evolution mechanism for this voltage oscillation phenomenon.}
	\item {A common Lyapunov function-based sufficient stability criterion for switched systems is proposed and proved. The stability conditions, solving procedures, and stability index calculation methods for the repeated LVRT process are derived by combining the WT-GSC switched system model and this stability criterion.}
	\item {The Sobol' global sensitivity analysis method is employed to identify the dominant parameters influencing system stability. Then, these parameters are optimized to maximize the stability index via the particle swarm optimization (PSO) algorithm.}
\end{enumerate}

The rest of this paper is organized as follows. Section II constructs the WT-GSC switched system model for the LVRT switching process. Section III analyzes the evolution mechanism of the wind turbine grid-connected point voltage oscillation. Section IV proposes the stability criterion and index for the WT-GSC switched system. Section V presents the sensitivity analysis and optimization methods for system parameters. Section VI verifies the proposed theories and results through simulation. Finally, conclusions are provided in Section VII.

\section{Switched System Model of the Wind Turbine Grid-Side Converter}
This section commences by introducing fundamental concepts and the generic mathematical model of switched systems. Then, taking the wind turbine's external connection impedance and internal control dynamics into account, dynamic equations describing its behaviors under normal operation and LVRT conditions are derived on the basis of the structure and control strategies of the WT-GSC. Finally, by incorporating a switching law that governs the switching between the aforementioned operational states, the WT-GSC switched system model for the repeated LVRT process is established.

\subsection{Fundamental Concepts of Switched Systems}
\begin{figure}[!htbp]
	\centering
	\includegraphics[width=9cm]{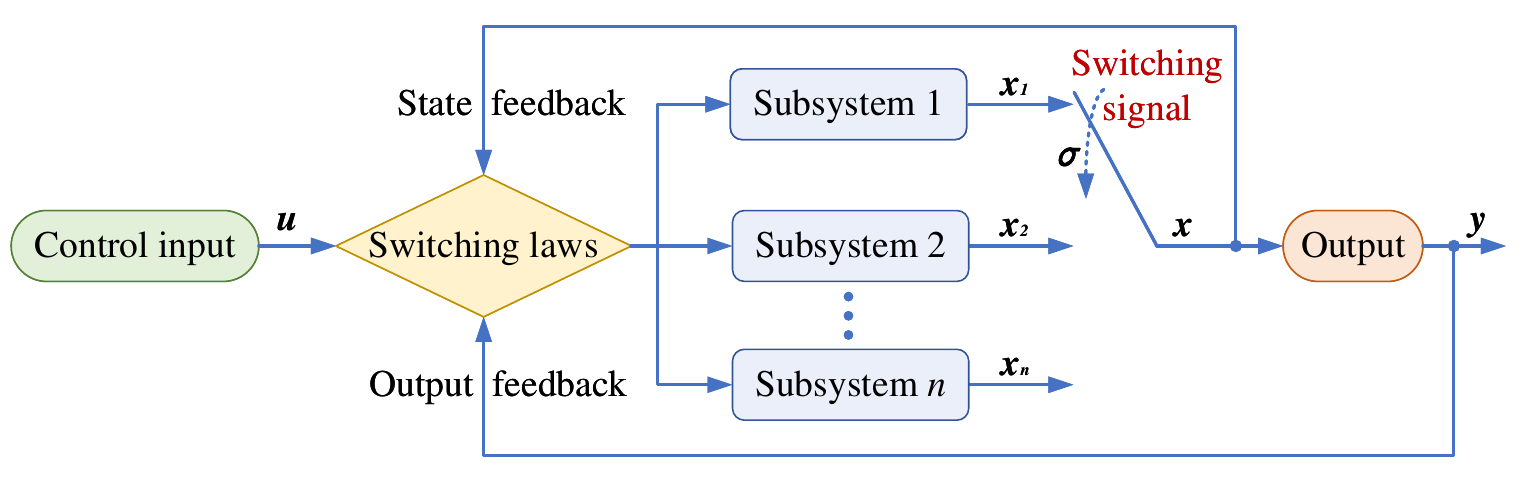}
	\caption{Diagram of a typical switched system.}
	\label{fig_1}
\end{figure}

The rapid advancement of modern computer control technology has facilitated widespread applications of switching control techniques across various fields, including automatic control, automotive industries, power systems, and many others \cite{Ref34_GaoJunweiTheoryandApplication2003, Ref35_ChenApplicationSwitchedSystem2014}. As depicted in Fig. \ref{fig_1}, a typical switched system contains a collection of continuous-time subsystems and associated switching laws orchestrating the switching among these subsystems. The dynamic evolution process of the switched system is determined by the dynamic of each subsystem and the corresponding switching laws.

The generic mathematical model of switched systems can be expressed as
\begin{equation} \label{eqn_1}
	\dot{\bm{x}} = {\bm{f}_\sigma}\left(\bm{x}\right), \ \sigma \in S
\end{equation}
where $\bm{x} \in \mathbb{R}^N$ is the continuous system state vector with respect to time, ${\bm{f}_\sigma}$ is a set of regular functions representing the dynamic equations of subsystems, $S = \{1,2,\cdots,n\}$ is an index set denoting the discrete states of the system with $n$ subsystems, and $\sigma:[0,\infty) \to S$ is a piecewise constant function of time taking values in $S$, which represents the switching signal. The value of $\sigma (t)$ at a given time $t$ depends on time or state vector values, or both, or more complex feedback effects \cite{Ref23_D.LiberzonBasicProblemsStability1999}.

If each individual subsystem is linear, the state space equation of the switched linear system can be expressed as
\begin{equation} \label{eqn_2}
	\begin{cases}
		\dot{\bm{x}} \left(t\right) = {\bm{A}_\sigma} \bm{x}\left(t\right) + {\bm{B}_\sigma} \bm{u}\left(t\right) \\[1mm]
		\bm{y} \left(t\right) = {\bm{C}_\sigma} \bm{x}\left(t\right) + {\bm{D}_\sigma} \bm{u}\left(t\right)
	\end{cases}
\end{equation}
where $\bm{u}$ is the control input vector, $\bm{y}$ is the output vector, and $\bm{A}_\sigma$, $\bm{B}_\sigma$, $\bm{C}_\sigma$, $\bm{D}_\sigma$ are constant matrices with appropriate dimensions when the value of $\sigma$ maintains the same \cite{Ref21_DCDC_switched_linear_systems_HuZongbo_2005}.

\subsection{Structure and Control Strategy of the WT-GSC}
The grid-side converter of the wind turbine comprises a controlled inverter bridge composed of insulated-gate bipolar transistors (IGBTs). It adopts a dual-loop control structure with a voltage outer-loop and a current inner-loop. The primary control objectives involve ensuring a consistent DC-side capacitor voltage level and generating reactive power in accordance with the predetermined reactive power reference value \cite{Ref39_WTG_equivalent_LiDongsheng_2024}.

Assuming the WT-GSC is connected to the grid through a filter with impedance ${R_g}+j\omega{L_g}$, and the connected point is the wind turbine's grid-connected point, the voltage equations of WT-GSC under the \textit{dq} coordinate system are obtained as
\begin{equation} \label{eqn_3}
	\begin{cases}
		{v_d} = - {R_g}{i_d} - {L_g}\dfrac{{\dif {i_d}}}{{\dif t}} + \omega{L_g}{i_q} + {v_{gd}} \\[2mm]
		{v_q} = - {R_g}{i_q} - {L_g}\dfrac{{\dif {i_q}}}{{\dif t}} - \omega{L_g}{i_d} + {v_{gq}}
	\end{cases}
\end{equation}
where $\omega$ is the synchronous electrical angular speed of the grid; $R_g$ and $L_g$ are the resistance and inductance of the filter, respectively; $v_d$ and $v_q$ are the \textit{d}- and \textit{q}-axes components of the WT-GSC AC-side voltage, respectively; $i_d$ and $i_q$ are the \textit{d}- and \textit{q}-axes components of the WT-GSC AC-side current, respectively;  $v_{gd}$ and $v_{gq}$ are the \textit{d}- and \textit{q}-axes components of the grid-connected point voltage, respectively.

The WT-GSC generally employs the grid-voltage-oriented vector control technique, which means the \textit{d}-axis of the synchronous rotating coordinate system is oriented in the direction of the grid-connected point voltage vector. Thus, we have ${v_{gd}} = {v_g}$ and ${v_{gq}} = 0$.

The power output equations at the grid-connected point of the WT-GSC are derived as
\begin{equation} \label{eqn_5}
	\begin{cases}
		{P_g} = 1.5\left( {{v_{gd}}{i_{gd}} + {v_{gq}}{i_{gq}}} \right) = 1.5{v_g}{i_d} \\[2mm]
		{Q_g} = 1.5\left( {{v_{gq}}{i_{gd}} - {v_{gd}}{i_{gq}}} \right) = - 1.5{v_g}{i_q}
	\end{cases}
\end{equation}
where $v_g$ is the grid-connected point voltage; $P_g$ and $Q_g$ are the active and reactive power output, respectively; $i_{gd}$ and $i_{gq}$ are the \textit{d}- and \textit{q}-axes components of the grid-connected current, respectively. Obviously, we have ${i_{gd}} = {i_d}$ and ${i_{gq}} = {i_q}$.

According to \eqref{eqn_5}, the decoupling control of active and reactive power for the WT-GSC is realized through the grid-voltage-oriented vector control technique. Hence, the \textit{d}- and \textit{q}-axes currents can be referred to as the active and reactive currents, respectively.

To summarize, the control equations for the voltage outer-loop and current inner-loop of the WT-GSC are expressed as
\begin{equation} \label{eqn_6}
	\begin{cases}
		{i_{d{\mathrm{ref}}}} = {K_{pDC}} ({V_{DC{\mathrm{ref}}}}\!-\!{V_{DC}}) + {K_{iDC}} \!\int\! {({V_{DC{\mathrm{ref}}}}\!-\!{V_{DC}})\dif t} \\[1mm]
		{i_{q{\mathrm{ref}}}} = {K_{pQ}} ({Q_{g{\mathrm{ref}}}}\!-\!{Q_g}) + {K_{iQ}} \!\int\! {({Q_{g{\mathrm{ref}}}}\!-\!{Q_g})\dif t}
	\end{cases}
\end{equation}
\begin{equation} \label{eqn_7}
	\begin{cases}
		{v_{d{\mathrm{ref}}}} = {K_{pd}} ({i_{d{\mathrm{ref}}}}\!-\!{i_d}) + {K_{id}} \!\int\! {({i_{d{\mathrm{ref}}}}\!-\!{i_d})\dif t} + \omega{L_g}{i_q} + {v_{gd}} \\[1mm]
		{v_{q{\mathrm{ref}}}} = {K_{pq}} ({i_{q{\mathrm{ref}}}}\!-\!{i_q}) + {K_{iq}} \!\int\! {({i_{q{\mathrm{ref}}}}\!-\!{i_q})\dif t} - \omega{L_g}{i_d} + {v_{gq}}
	\end{cases}
\end{equation}
where $i_{d{\mathrm{ref}}}$ and $i_{q{\mathrm{ref}}}$ are the \textit{d}- and \textit{q}-axes current reference values from the voltage loop, respectively; $v_{d{\mathrm{ref}}}$ and $v_{q{\mathrm{ref}}}$ are the \textit{d}- and \textit{q}-axes voltage reference values from the current loop, respectively; $V_{DC}$ and $V_{DC{\mathrm{ref}}}$ are the actual and reference values of the DC-side voltage, respectively; $Q_g$ and $Q_{g{\mathrm{ref}}}$ are the actual and reference values of the reactive power output, respectively; $K_{pDC}$ and $K_{iDC}$ are the PI parameters for fixed DC-side voltage control; $K_{pQ}$ and $K_{iQ}$ are the PI parameters for fixed reactive power control; $K_{pd}$, $K_{id}$, $K_{pq}$, and $K_{iq}$ are the PI parameters for current inner-loop control.



\subsection{Construction of the WT-GSC Switched System Model}
Assuming the voltage pulse width modulation (PWM) control response of the WT-GSC controlled inverter bridge is fast enough so that the AC-side voltage quickly tracks its reference value, we have ${v_{d{\mathrm{ref}}}} \approx {v_d}$ and ${v_{q{\mathrm{ref}}}} \approx {v_q}$. Combining this assumption with \eqref{eqn_7} approximately yields
\begin{equation} \label{eqn_8}
	\!\!\!\!\begin{cases}
		\!{v_d} \!-\! {v_{gd}} \!=\! {K_{pd}} ({i_{d{\mathrm{ref}}}}\!-\!{i_d}) \!+\!\! {K_{id}} \!\int\!\! {({i_{d{\mathrm{ref}}}}\!-\!{i_d}) \! \dif t \!+\! \omega{L_g}{i_q}} \\[1mm]
		\!{v_q} \!-\! {v_{gq}} \!=\! {K_{pq}} ({i_{q{\mathrm{ref}}}}\!-\!{i_q}) \!+\!\! {K_{iq}} \!\int\!\! {({i_{q{\mathrm{ref}}}}\!-\!{i_q}) \! \dif t \!-\! \omega{L_g}{i_d}}.
	\end{cases}
\end{equation}

Generally, the difference values between the AC-side voltage and the grid-connected point voltage (i.e., ${v_d}-{v_{gd}}$ and ${v_q}-{v_{gq}}$) fluctuate very little and can be approximated as constants. Thus, taking the time derivatives of both sides of \eqref{eqn_8} gives
\begin{equation} \label{eqn_9}
	\begin{cases}
		\! {K_{pd}} \! \left(\dfrac{{\dif {i_{d{\mathrm{ref}}}}}}{{\dif t}} \!-\! \dfrac{{\dif {i_d}}}{{\dif t}}\right) \!+\! {K_{id}} \left({i_{d{\mathrm{ref}}}} \!-\! {i_d}\right) \!+\! \omega{L_g}\dfrac{{\dif {i_q}}}{{\dif t}} \!=\! 0 \\[3mm]
		\! {K_{pq}} \! \left(\dfrac{{\dif {i_{q{\mathrm{ref}}}}}}{{\dif t}} \!-\! \dfrac{{\dif {i_q}}}{{\dif t}}\right) \!+\! {K_{iq}} \left({i_{q{\mathrm{ref}}}} \!-\! {i_q}\right) \!-\! \omega{L_g}\dfrac{{\dif {i_d}}}{{\dif t}} \!=\! 0.
	\end{cases}
\end{equation}

Considering different reactive power control strategies under the normal operation and LVRT conditions of the wind turbine, the state equations of corresponding subsystems can be derived on the basis of \eqref{eqn_9}.

\subsubsection{Normal Operation Subsystem}
When the wind turbine works under the normal operation condition, ${i_{d{\mathrm{ref}}}}$ is determined by fixed DC-side voltage control to maintain a constant capacitor voltage, and ${i_{q{\mathrm{ref}}}}$ is determined by fixed reactive power control, with the reactive power reference value usually set to zero to achieve unit power factor operation. Thus, we obtain ${i_{d{\mathrm{ref}}}} = \mathrm{const}$ and ${i_{q{\mathrm{ref}}}} = 0$.
Substituting this result into \eqref{eqn_9} and then reorganizing the equation yields the state equation of the normal operation subsystem, which is expressed as
\begin{equation} \label{eqn_11}
	\begin{aligned}
		& \dot{\bm{x}} = {\bm{A}_1}\bm{x}+{\bm{B}_1}{\bm{u}_1}, \ \sigma = 1 \\[1mm]
		& {\bm{A}_1} = \dfrac{1}{{{\omega ^2}L_g^2+{K_{pd}}{K_{pq}}}} {\begin{bmatrix}
			{-{K_{pq}}{K_{id}}} & {-\omega{L_g}{K_{iq}}} \\[1mm]
			{\omega {L_g}{K_{id}}} & {-{K_{pd}}{K_{iq}}}
		\end{bmatrix}} \\[1mm]
		& {\bm{B}_1} = \dfrac{1}{{{\omega ^2}L_g^2+{K_{pd}}{K_{pq}}}} {\begin{bmatrix}
			{{K_{pq}}{K_{id}}} & {\omega{L_g}{K_{iq}}} \\[1mm]
			{-\omega{L_g}{K_{id}}} & {{K_{pd}}{K_{iq}}}
		\end{bmatrix}}
	\end{aligned}
\end{equation}
where $\bm{x}={[{i_d}, \ {i_q}]^\T}$, ${\bm{u}_1}={[{I_{d1}}, \ 0]^\T}$, $I_{d1}$ is a constant, and $\sigma$ is the discrete state variable.

\subsubsection{LVRT Subsystem}
According to \textit{Technical specification for connecting wind farm to power system} of China, the reactive power priority control strategy is adopted when the wind turbine works under the LVRT condition \cite{Ref15_GBT_wind_farm_specification_2021}. This strategy prioritizes the reactive current requirements, then the reference value of active current is determined by fixed DC-side voltage control and the current constraints of the WT-GSC, which is denoted as
\begin{equation} \label{eqn_12}
	\begin{cases}
		{i_{d{\mathrm{ref}}}} \!=\! \min \{ {{I_{d2}}}, \ {{i_{d\max }}} \}, \ {i_{d\max}} \!=\! \sqrt {I_{\max}^2 - i_{q{\mathrm{ref}}}^2} \\[1mm]
		{i_{q{\mathrm{ref}}}} \!=\! {K_1}(0.9 - v_g ), 0.2 \le \!v_g\! \le 0.9, 1.5 \le \!{K_1}\! \le 3
	\end{cases}
\end{equation}
where $K_1$ is the dynamic reactive current proportional coefficient, $I_{d2}$ is the constant active current reference value given by fixed DC-side voltage control, $i_{d\max}$ is the upper limit of active current, and $I_{\max}$ is the upper limit of WT-GSC current.

Supposing the wind turbine is connected to the grid through a transmission line with impedance $R+j \omega L$, the grid voltage is a constant $v_G$, and the capacity of WT-GSC is large enough, then we combine \eqref{eqn_5} and power flow equations to obtain the grid-connected point voltage as
\begin{equation} \label{eqn_13}
	\begin{aligned}
		v_g & \approx v_G + \Delta{v} = v_G  + \dfrac{{{P_g}R+{Q_g}X}} {{{v_g}}} \\
		&= v_G + 1.5R{i_d} - 1.5 \omega L i_q.
	\end{aligned}
\end{equation}

Substituting \eqref{eqn_13} into \eqref{eqn_12} and taking the time derivative yields
%
\begin{equation} \label{eqn_15}
	\dfrac{{\dif {i_{q{\mathrm{ref}}}}}}{{\dif t}} = {K_1} \left(-1.5R\dfrac{{\dif {i_d}}}{{\dif t}} + 1.5 \omega L \dfrac{{\dif {i_q}}}{{\dif t}}\right).
\end{equation}

By substituting \eqref{eqn_11}$-$\eqref{eqn_15} into \eqref{eqn_9} and reorganizing, the state equation of the LVRT subsystem is expressed as
\begin{equation} \label{eqn_16}
	\begin{aligned}
		& \dot{\bm{x}} = {\bm{A}_2}\bm{x}+{\bm{B}_2}{\bm{u}_2}, \ \sigma = 2 \\[1mm]
		& {\bm{A}_2} = {(\bm{I}-\bm{E} \times \bm{F})^{-1}} \times ({\bm{A}_1}+{\bm{B}_1} \times \bm{F}) \\[1mm]
		& {\bm{B}_2} = {(\bm{I}-\bm{E} \times \bm{F})^{-1}} \times {\bm{B}_1} \\[1mm]
		& \bm{E} = \dfrac{1}{{{\omega^2}L_g^2+{K_{pd}}{K_{pq}}}} {\begin{bmatrix}
			{{K_{pd}}{K_{pq}}} & {\omega{L_g}{K_{pq}}} \\[1mm]
			{-\omega{L_g}{K_{pd}}} & {{K_{pd}}{K_{pq}}}
		\end{bmatrix}} \\[1mm]
		& \bm{F} = 1.5 K_1 {\begin{bmatrix}
			0 & 0 \\[1mm]
			{-R} & {\omega L}
		\end{bmatrix}}
	\end{aligned}
\end{equation}
where $\bm{I}$ is the identity matrix, $\bm{E}$ and $\bm{F}$ are intermediate matrices, ${\bm{u}_2}={[{I_{d2}}, \ {K_1}(0.9-v_G)]^\T}$, and $I_{d2}$ is a constant.

In conclusion, by summarizing \eqref{eqn_11} and \eqref{eqn_16}, the WT-GSC switched system model for the repeated LVRT process is obtained as
\begin{equation} \label{eqn_17}
	\begin{cases}
		\dot{\bm{x}} = {\bm{A}_\sigma}\bm{x}+{\bm{B}_\sigma}{\bm{u}_\sigma}, \ \sigma \in \{1,2\} \\[1mm]
		\{\bm{A}_1, \bm{B}_1, \bm{u}_1\} \xrightleftharpoons[v_g \ge v_{\mathrm{LVRT}}]{v_g < v_{\mathrm{LVRT}}} \{\bm{A}_2, \bm{B}_2, \bm{u}_2\}
	\end{cases}
\end{equation}
where $v_{{\mathrm{LVRT}}}$ denotes the LVRT threshold of the wind turbine.

\section{Mechanism Analysis of the Grid-Connected Point Voltage Oscillation}

Based on the WT-GSC switched system model combined with the control parameters and connection impedance values shown in Table \ref{table_1}, this section further explores the switching dynamic process of repeated LVRT and elucidates the underlying mechanism of voltage oscillations.

\begin{table}[!htbp]
	\caption{Control Parameters and Connection Impedance}
	\setstretch{0.9}
	\label{table_1}
	\centering
	\resizebox{\columnwidth}{!}{
	\begin{tabular}{ccc}
		\toprule
		\toprule
		Symbol & Description & Value (p.u.) \\
		\midrule
		$K_{pd}$, $K_{pq}$ & Current inner-loop proportional coefficient & 0.10 \\[1mm]
		$K_{id}$, $K_{iq}$ & Current inner-loop integral coefficient & 5.00 \\[1mm]
		$K_1$ & Dynamic reactive current coefficient & 2.00 \\[1mm]
		$L_g$ & Filter inductance & 3.25e-4 \\[1mm]
		$R$   & Line resistance   & 7.58e-4 \\[1mm]
		$L$   & Line inductance   & 1.00e-3 \\
		\bottomrule
		\bottomrule
	\end{tabular}
	}
\end{table}

\subsection{Evolution Dynamic and Mechanism of Voltage Oscillations}
\begin{figure}[!htbp]
	\centering
	\vspace{-3mm}
	\subfloat[]{
		\includegraphics[width=4.2cm]{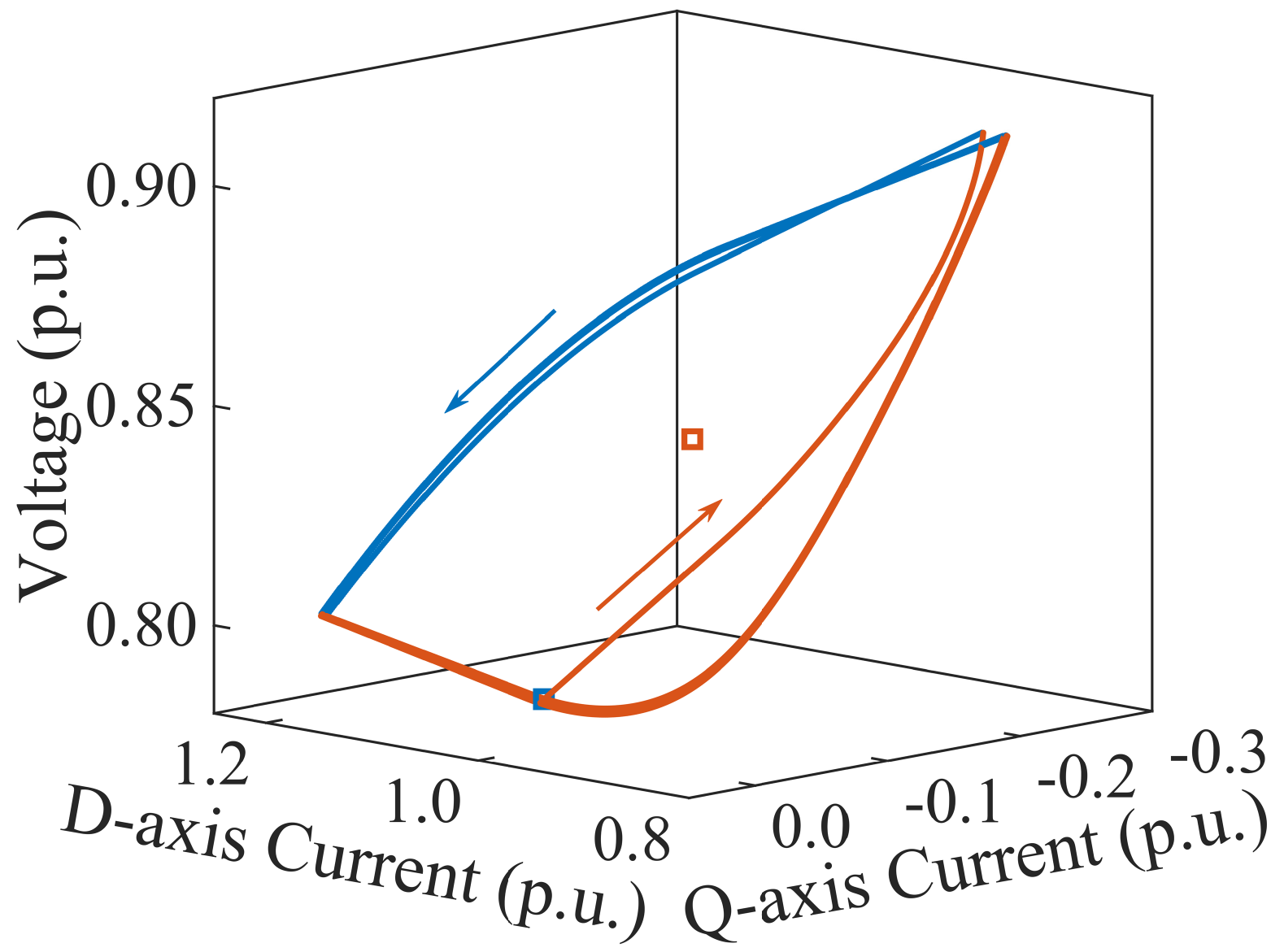}
		\label{fig_2a}
	}
	\subfloat[]{
		\includegraphics[width=4.2cm]{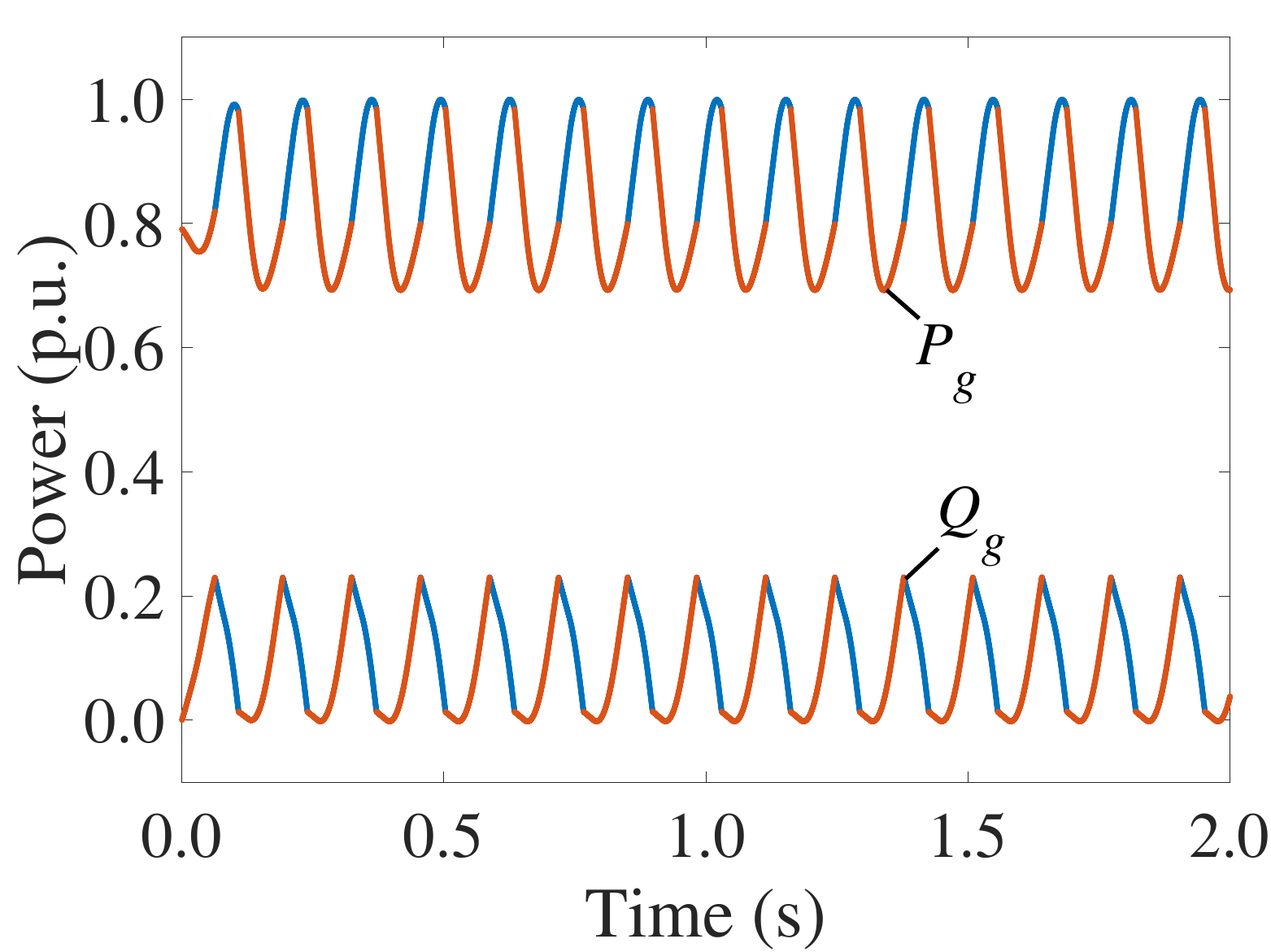}
		\label{fig_2b}
	}
	\caption{Dynamic evolution curves of the WT-GSC switched system during the repeated LVRT process. (a) Phase portrait. (b) Power curve. The blue and orange curves denote the normal operation and LVRT subsystem dynamics, respectively. The blue and orange square points in (a) denote the stable equilibrium points of the normal operation and LVRT subsystems, respectively (i.e., (1.00, 0, 0.79) and (1.00, -0.11, 0.84)).}
	\label{fig_2}
\end{figure}

When the wind turbine is integrated into a weak power grid and generates large amounts of active power, repeated LVRT may occur, leading to problems such as voltage oscillations and overvoltages at the grid-connected point. The essence of this phenomenon lies in the wind turbine repeatedly switching between the normal operation and LVRT subsystems. By substituting the parameters in Table \ref{table_1} into the WT-GSC switched system model \eqref{eqn_17} and setting the initial point of system trajectory to the stable equilibrium point of the normal operation subsystem, the phase portrait and output power curves of the switched system during the repeated LVRT process are shown in Fig. \ref{fig_2}.

As can be seen from Fig. \ref{fig_2}, during normal operation, the \textit{d}-axis current and active power increase while the \textit{q}-axis current and reactive power decrease, until the voltage drops below $v_{{\mathrm{LVRT}}}$. After entering LVRT, the \textit{d}-axis current and active power decrease while the \textit{q}-axis current and reactive power increase, until the voltage is higher than $v_{{\mathrm{LVRT}}}$. Moreover, the excessive reactive power compensation from the LVRT subsystem can induce overvoltages. If the time delay of withdrawing dynamic reactive current increments is long, the phase trajectory of the switched system will evolve with the LVRT subsystem for an extended period, resulting in overvoltages and subsequent voltage oscillations.

\subsection{System Trajectories Under Different Working Conditions}
\begin{figure}[!htbp]
	\centering
	\vspace{-3mm}
	\subfloat[]{
		\includegraphics[width=4.2cm]{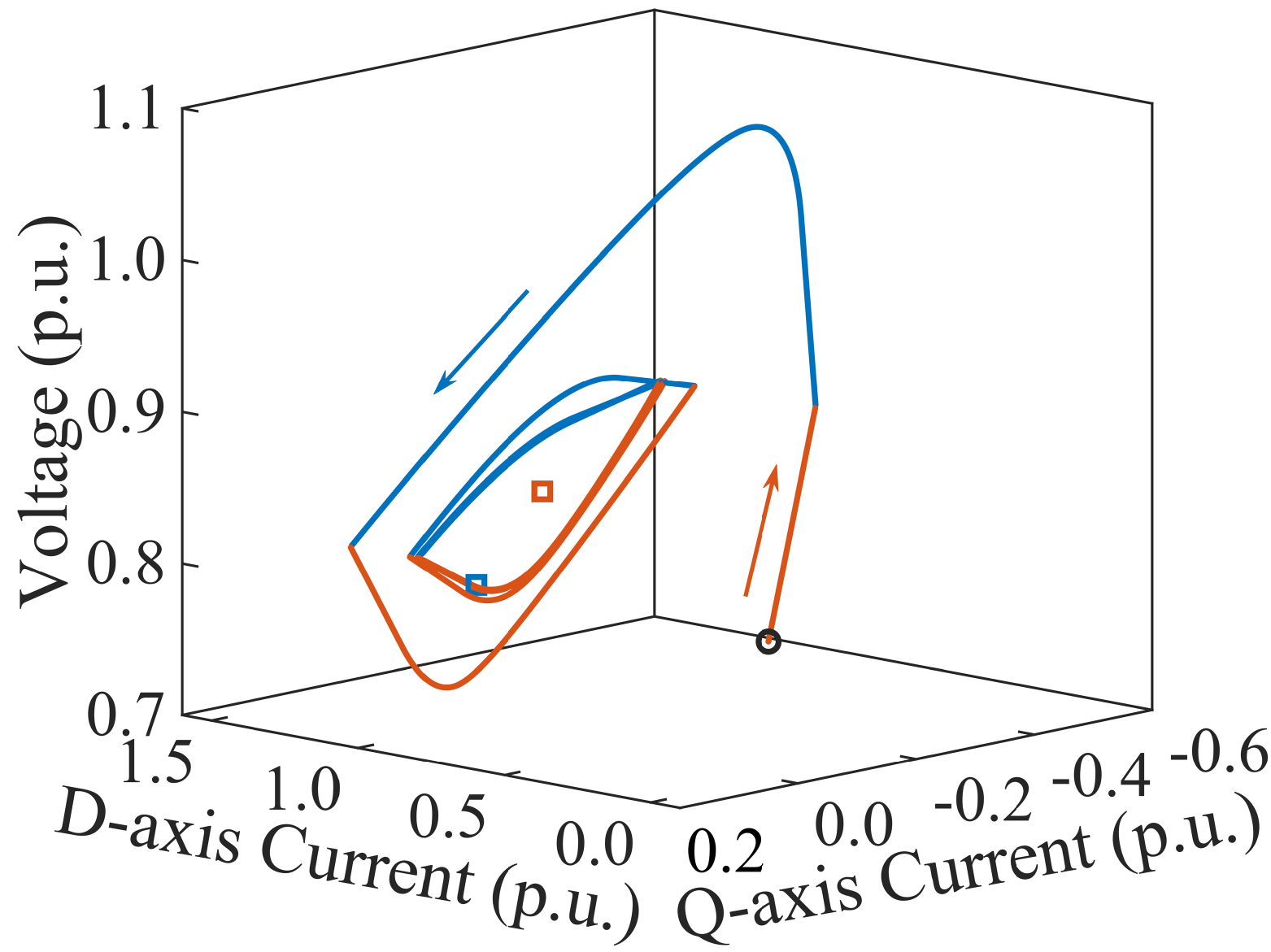}
		\label{fig_3a}
	}
	\subfloat[]{
		\includegraphics[width=4.2cm]{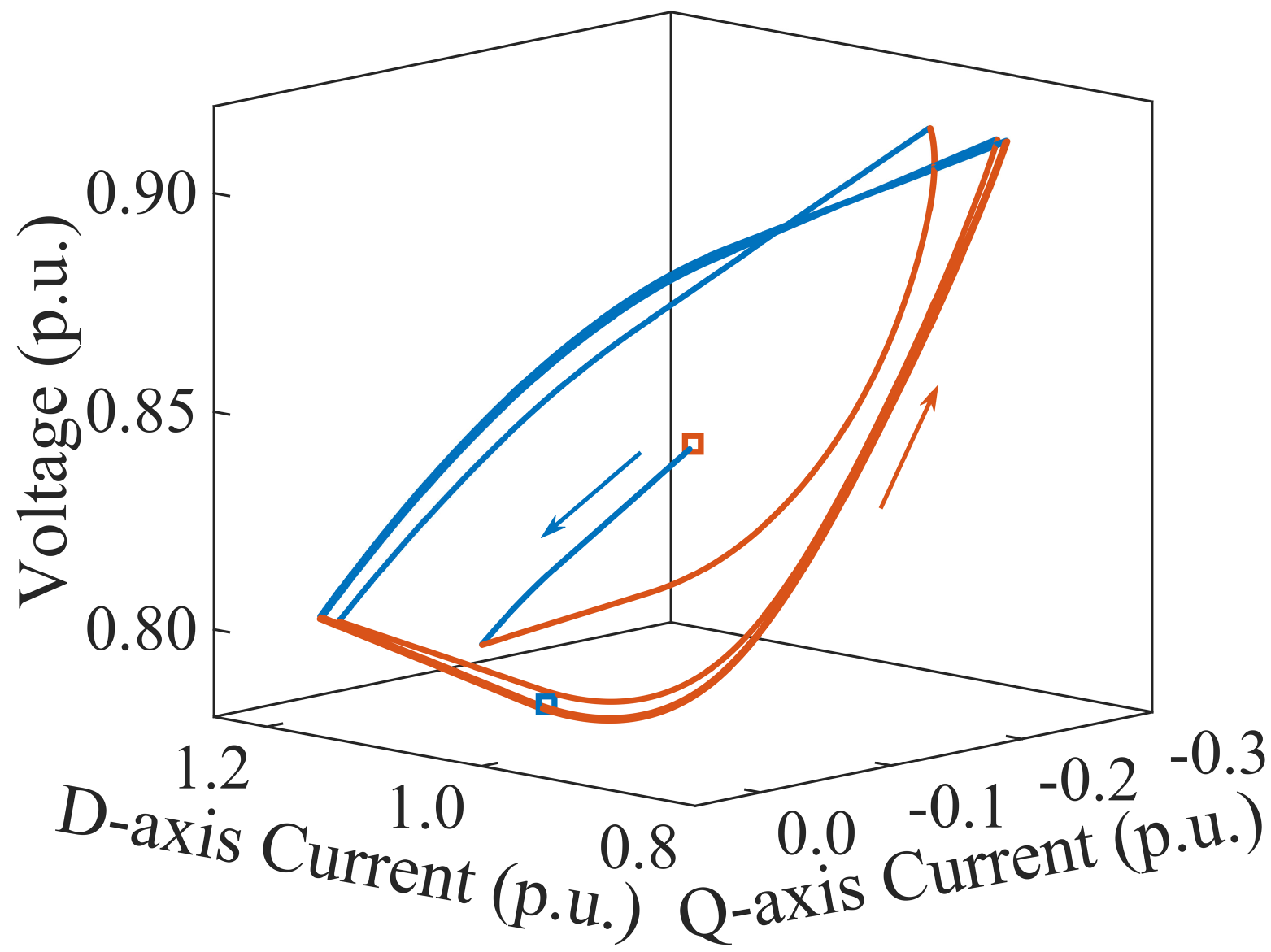}
		\label{fig_3b}
	}
	\caption{Phase portraits of the WT-GSC switched system starting from different initial points. (a) Starting from the zero-state point (0, 0, 0.79). (b) Starting from the stable equilibrium point of the LVRT subsystem (1.00, -0.11, 0.84).}
	\label{fig_3}
\end{figure}

To further validate the consistency of conclusions obtained with different initial points of system trajectories, the initial point is set to the system zero-state point and the stable equilibrium point of the LVRT subsystem, respectively. The system phase portraits are shown in Fig. \ref{fig_3}.

As shown in Fig. \ref{fig_2a} and Fig. \ref{fig_3}, due to the reactive power switching control strategy, the stable equilibrium points of the normal operation and LVRT subsystems are no longer stable for the switched system. Under this parameter condition, the system trajectories starting from any points within the current constraints will ultimately undergo repeated LVRT, resulting in voltage oscillations. This proves that the voltage oscillation phenomenon originates from the difference between the stable equilibrium points of these two subsystems and is independent of the initial points.

\begin{figure}[!htbp]
	\centering
	\vspace{-3mm}
	\subfloat[]{
		\includegraphics[width=4.2cm]{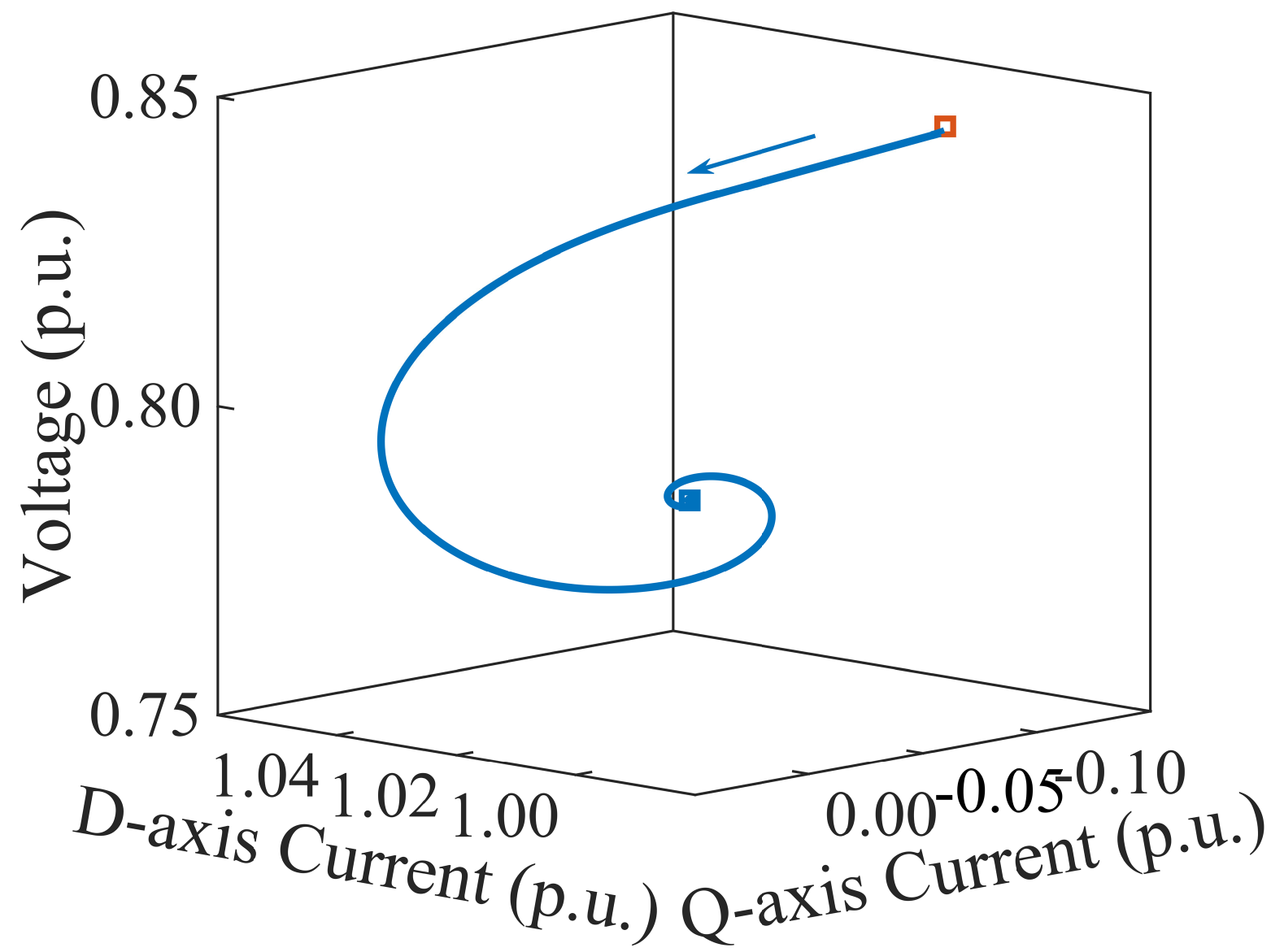}
		\label{fig_4a}
	}
	\subfloat[]{
		\includegraphics[width=4.2cm]{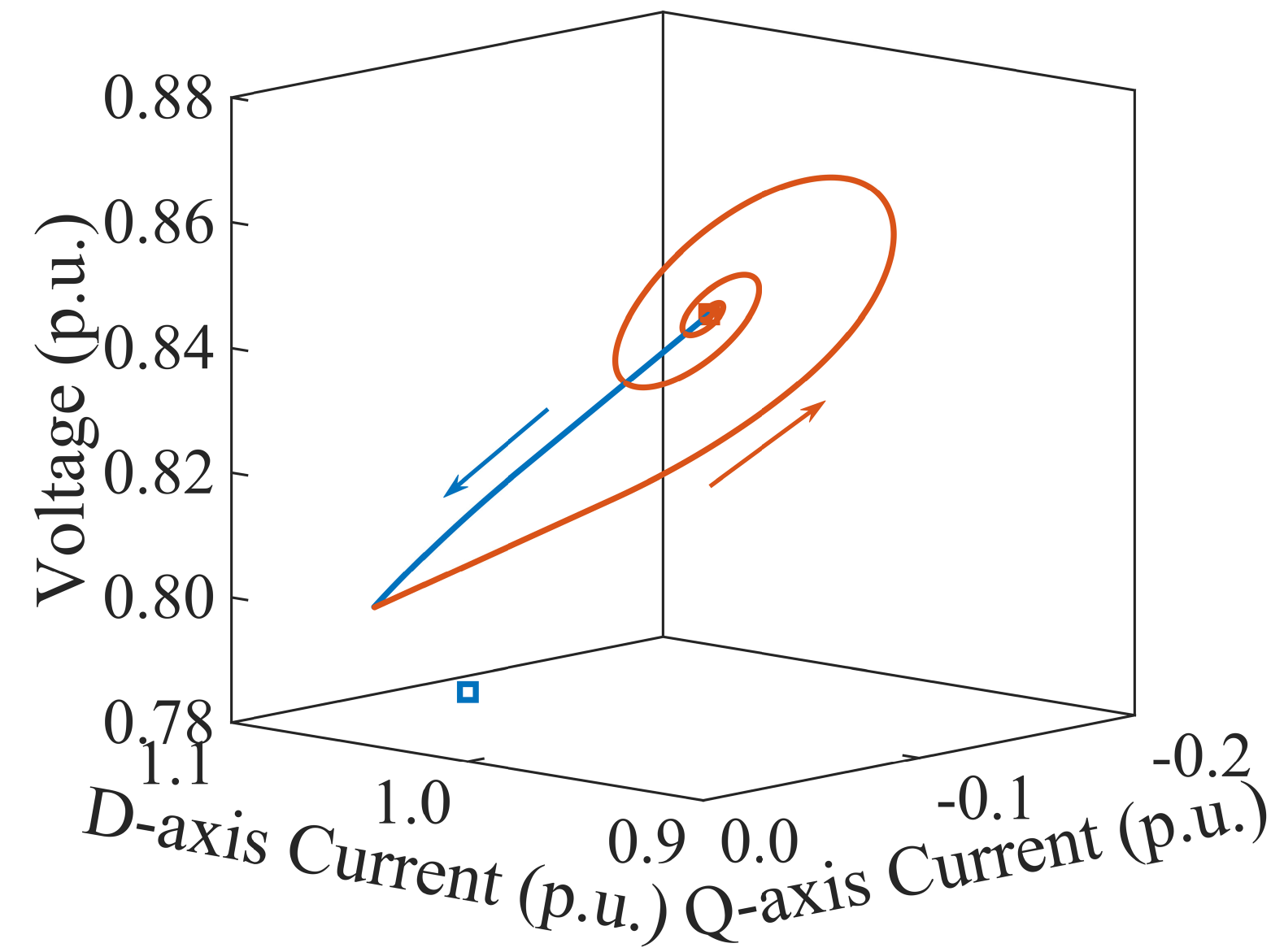}
		\label{fig_4b}
	}
	\caption{Phase portraits of the WT-GSC switched system under different working conditions. (a) Blocked LVRT control. (b) Continuous LVRT operation.}
	\label{fig_4}
\end{figure}

In order to illustrate that repeated LVRT of the wind turbine is the inducement of voltage oscillations, the LVRT control is blocked under identical parameters to obtain the system phase portrait as shown in Fig. \ref{fig_4a}. Moreover, when $K_{pd}$ and $K_{pq}$ are changed to 0.15 with other parameters unchanged, the continuous LVRT dynamic process of the WT-GSC switched system is obtained, which is shown in Fig. \ref{fig_4b}.

As shown in Fig. \ref{fig_4a}, when the LVRT control is blocked, the system eventually stabilizes at the normal operation subsystem's equilibrium point, where the voltage value is below 0.80p.u., exceeding the normal operation boundary. This is a virtual equilibrium point of the WT-GSC switched system \cite{BernardoBifurcationsNonsmoothDynamical2008}, which will not be reached under the switching control. In Fig. \ref{fig_4b}, the LVRT subsystem's equilibrium point is within the operation boundary of this subsystem, which is a regular equilibrium point of the switched system. When the dynamic reactive current increment supplied during the LVRT condition is insufficient or large LVRT switching hysteresis is adopted, the switched system will stabilize at this point, causing continuous LVRT operation.

In summary, due to the different stable equilibrium points of the normal operation and LVRT subsystems, the system trajectories keep switching near the LVRT threshold under the reactive power switching control. This switching process makes the wind turbine repeatedly enter and exit LVRT, thus inducing the grid-connected point voltage oscillation.

\section{Stability Analysis and Assessment of the WT-GSC Switched System}
A switched system consists of multiple subsystems and switching laws that determine the switching mechanism among them. Consequently, the overall stability of the switched system depends on the stability of each subsystem and the defined switching laws. Even if all subsystems are stable/unstable, the switched system is not necessarily stable/unstable. Therefore, this section proposes quantitative methods for comprehensive stability analysis and assessment of the WT-GSC switched system.

\subsection{Definition and Proof of Switched System Stability}
{ 
	\newtheoremstyle{definition_me}{3pt}{3pt}{\normalshape}{0cm}{\bfseries}{:}{0.2em}{}
	\theoremstyle{definition_me}
	\newtheorem{definition_me}{\indent Definition}
}
\begin{definition_me}[\textbf{Switched System Stability \cite{Ref19_switching_in_systems_Liberzon_2003}}]
	For the switched system in \eqref{eqn_1}, if there exist a positive constant $\delta$ and a function $\beta \left( r,t \right):\left[ {0,\infty } \right) \!\times\! \left[ {0,\infty} \right) \!\to\! \left[ {0,\infty } \right)$, such that for all switching signals $\sigma$, the solutions of the switched system with $\Vert {\bm{x}(0)} \Vert \le \delta$ satisfy the inequality:
	\begin{equation} \label{eqn_18}
		\Vert {\bm{x}(t)} \Vert \le \beta \left( {\Vert {\bm{x}(0)} \Vert,t} \right), \ \forall t \ge 0
	\end{equation}
	where $\beta \left( r,t \right)$ is a strictly monotonic increasing continuous function with $\beta \left( {0,t} \right)=0$ for each fixed $t \ge 0$, and $\beta \left( r,t \right)$ decreases to zero as $t \!\to\! \infty$ for each fixed $r \ge 0$. Then, the switched system is asymptotically stable. If this inequality holds for all switching signals and initial conditions, the switched system is globally asymptotically stable.
\end{definition_me}

\begin{definition_me}[\textbf{Common Lyapunov Function \cite{Ref19_switching_in_systems_Liberzon_2003}}]
	For the switched system in \eqref{eqn_1}, given a positive definite continuously differentiable function $V (\bm{x}):{\mathbb{R}^N} \to \mathbb{R}$, if there exists a positive definite continuous function $W (\bm{x}):{\mathbb{R}^N} \to \mathbb{R}$ such that
	\begin{equation} \label{eqn_19}
		\dfrac{{\partial V}}{{\partial \bm{x}^\T}}{\bm{f}_\sigma }\left( \bm{x} \right) \le - W\left( \bm{x} \right), \ \forall \sigma \in S
	\end{equation}
	then the function $V\left(\bm{x}\right)$ is called the common Lyapunov function of the switched system.
\end{definition_me}

The matrix convex combinations of subsystems in the switched system are defined as
\begin{equation} \label{eqn_20}
	\begin{cases}
		\! {\bm{A}_{\bm{\lambda}}} = \sum\limits_{i=1}^n {{\lambda_i}{\bm{A}_i}}, \quad {\bm{B}_{\bm{\lambda}}}{\bm{u}_{\bm{\lambda}}} = \sum\limits_{i=1}^n {{\lambda_i}{\bm{B}_i}{\bm{u}_i}} \\
		\! \bm{\Lambda} \!=\! \{ \bm{\lambda} \!=\! [{\lambda_1},{\lambda_2},\!\cdots\!,{\lambda_n}] \!\in\! {\mathbb{R}^n},{\lambda_i} \ge 0, \sum\limits_{i=1}^n \! {{\lambda_i}} \!=\! 1 \}
	\end{cases}
\end{equation}
where ${\lambda _i}$ can be interpreted as the proportion of subsystem $i$ on the switched system solution trajectory.

{
	\newtheoremstyle{theorem_me}{3pt}{3pt}{\normalshape}{0cm}{\bfseries}{:}{0.2em}{}
	\theoremstyle{theorem_me}
	\newtheorem{theorem_me}{\indent Theorem}
}
\begin{theorem_me}[\textbf{Stability Criterion \cite{Ref32_DeaectoSwitchedAffineSystems2010, Ref33_MaPeriodicTimeTriggeredHybrid2023}}]
	Consider the switched system described as
	\begin{equation} \label{eqn_21}
		\dot{\bm{x}} = {\bm{A}_\sigma}\bm{x} + {\bm{B}_\sigma}{\bm{u}_\sigma}, \ \sigma (t):t \ge 0 \to S = \{1,2,\cdots,n\}.
	\end{equation}
	For a specific equilibrium point $\bm{x}_e$, if there exist a symmetric positive definite matrix $\bm{P} \in {\mathbb{R}^{n \times n}}$, a constant $\mu > 0$, and a set of $\bm{\lambda} \in \bm{\Lambda} $ satisfying the conditions as follows:
	\begin{subequations} \label{eqn_22}
		\begin{align}
			& {\bm{A}_{\bm{\lambda}}}{\bm{x}_e}+{\bm{B}_{\bm{\lambda}}}{\bm{u}_{\bm{\lambda}}} = \bm{0} \label{eqn_22a} \\[1mm]
			& \bm{A}_i^\T \bm{P} + \bm{P}{\bm{A}_i} + \mu \bm{P} \prec \bm{0}, \ \forall i \in S \label{eqn_22b} \\[1mm]
			& \sigma = \arg \mathop {\min}\limits_{i \in S} {\bm{\xi}^\T} \bm{P} \bm{b}_i \label{eqn_22c}
		\end{align}
	\end{subequations}
	where $\bm{\xi} = \bm{x}-{\bm{x}_e}$, ${\bm{b}_i} = {\bm{A}_i}{\bm{x}_e} + {\bm{B}_i}{\bm{u}_i}$, and $\arg \mathop {\min}$ indicates to obtain the value of $\sigma $ as the value of $i$ that minimizes the results of ${\bm{\xi}^\T} \bm{P} \bm{b}_i$. Then, the switched system is asymptotically stable about $\bm{x}_e$.
\end{theorem_me}

\begin{proof}
	Adopt a quadratic common Lyapunov function $ V \left(\bm{\xi} \right) = {\bm{\xi}^\T} \bm{P} \bm{\xi}$, where $V\left(\bm{\xi} \right) \ge 0$, and $V\left(\bm{0}\right)=0$. Combining \eqref{eqn_20}$-$\eqref{eqn_22}, the time derivative of $V\left(\bm{\bm{\xi}}\right)$ is derived as
	\begin{equation} \label{eqn_25}
		\begin{aligned}
			\dot V\left(\bm{\xi}\right)
			&= {{\dot{\bm{x}}}^\T} \bm{P} \bm{\xi}  + {\bm{\xi} ^\T} \bm{P} \dot{\bm{x}} \\[0.5mm]
			&= 2{\bm{\xi}^\T} \bm{P} ({\bm{A}_{\sigma}}\bm{x} + {\bm{B}_{\sigma}}{\bm{u}_{\sigma}}) \\[0.5mm]
			&= 2{\bm{\xi}^\T} \bm{P} ({\bm{A}_{\sigma}}{\bm{x}_e} + {\bm{B}_{\sigma}}{\bm{u}_{\sigma}}) + {\bm{\xi}^\T}(\bm{A}_{\sigma}^\T \bm{P} + \bm{P} {\bm{A}_{\sigma}})\bm{\xi} \\[0.5mm]
			&= \mathop{\min} \limits_{i \in {\mathbb{N}^+}} 2{\bm{\xi} ^\T} \bm{P} ({\bm{A}_i}{\bm{x}_e} + {\bm{B}_i}{\bm{u}_i}) + {\bm{\xi} ^\T}(\bm{A}_\sigma ^\T \bm{P} + \bm{P}{\bm{A}_\sigma})\bm{\xi} \\[0.5mm]
			&< \mathop{\min} \limits_{\bm{\lambda} \in \bm{\Lambda}} 2{\bm{\xi} ^\T} \bm{P} ({\bm{A}_{\bm{\lambda}}}{\bm{x}_e} + {\bm{B}_{\bm{\lambda}}}{\bm{u}_{\bm{\lambda}}}) - \mu {\bm{\xi} ^\T} \bm{P} \bm{\xi} \\[0.5mm]
			&= - \mu {\bm{\xi} ^\T} \bm{P} \bm{\xi} < 0
		\end{aligned}
	\end{equation}
	which obtains $\dot V\left(\bm{\xi}\right)<0$. Therefore, according to the Lyapunov stability criterion, the switched system is asymptotically stable about $\bm{x}_e$, which completes the proof.
\end{proof}


For a switched system with $n$ subsystems, \eqref{eqn_22c} is equivalent to
\begin{equation} \label{eqn_24}
	\sigma = j, \quad \mathrm{if} \ {{\bm{\xi}^\T} \bm{P} \bm{b}_j} \le \mathop \forall \limits_{i \ne j} {{\bm{\xi}^\T} \bm{P} \bm{b}_i}
\end{equation}
where $1 \le i \ne j \le n$ and $n \ge 2$.

%
%

\subsection{Sufficient Stability Criterion of WT-GSC Switched System}

In this subsection, by combining the WT-GSC switched system model \eqref{eqn_17} and the stability analysis method from Theorem 1, a sufficient stability criterion for the repeated LVRT switching process of the wind turbine is obtained.

When the wind turbine works under normal operation, $v_g \ge v_{{\mathrm{LVRT}}}$ is satisfied, that is
\begin{equation} \label{eqn_29}
	- 1.5R{i_d} + 1.5 \omega L {i_q} + v_{{\mathrm{LVRT}}} - v_G \le 0.
\end{equation}

In order to find the connection between the stable switching law \eqref{eqn_24} and the LVRT criterion \eqref{eqn_29}, \eqref{eqn_24} is expanded as
\begin{equation} \label{eqn_30}
	{\bm{\xi}^\T} \bm{P} \left( \bm{b}_1 - \bm{b}_2 \right) = {\begin{bmatrix}
			{{x_1}\!-\!{x_{1e}}} \!&\! {{x_2}\!-\!{x_{2e}}}
	\end{bmatrix}}
	{\begin{bmatrix}
			{{p_{11}}} \!&\! {{p_{12}}} \\[1mm]
			{{p_{21}}} \!&\! {{p_{22}}}
	\end{bmatrix}}
	{\begin{bmatrix}
			{{d_1}} \\[1mm]
			{{d_2}}
	\end{bmatrix}}
\end{equation}
where $\bm{x} = {[{x_1}, \ {x_2}]^\T}$, $\bm{x}_e = {[{x_{1e}}, \ {x_{2e}}]^\T}$, and ${[{d_1}, \ {d_2}]^\T} = \bm{b}_1 - \bm{b}_2$.

Combining \eqref{eqn_29} and \eqref{eqn_30}, the state variables $x_1$ and $x_2$ correspond to $i_d$ and $i_q$, respectively. Moreover, $\bm{P}$ is a symmetric positive definite matrix with ${p_{12}} = {p_{21}}$. It is obtained by the undetermined coefficient method that
\begin{equation} \label{eqn_32}
		{\begin{bmatrix}
			{{d_1}}&{{d_2}}&0&0\\[1mm]
			0&0&{{d_1}}&{{d_2}}\\[1mm]
			0&1&{-1}&0\\[1mm]
			{{d_1}{x_{1e}}}&{{d_2}{x_{1e}}}&{{d_1}{x_{2e}}}&{{d_2}{x_{2e}}}
		\end{bmatrix}} \!
		{\begin{bmatrix}
			{{p_{11}}}\\[1mm]
			{{p_{12}}}\\[1mm]
			{{p_{21}}}\\[1mm]
			{{p_{22}}}
		\end{bmatrix}} \!=\! 
		{\begin{bmatrix}
			{-1.5R}\\[1mm]
			{1.5 \omega L}\\[1mm]
			0\\[1mm]
			{v_G \!-\! v_{{\mathrm{LVRT}}}}
		\end{bmatrix}}.
\end{equation}

When the state equations of these two subsystems are different, we have ${[{d_1}, \ {d_2}]^\T} \ne \textbf{0}$, then the rank of the coefficient matrix in \eqref{eqn_32} is 3. Let ${p_{11}}=p$, eliminating $p_{12}$, $p_{21}$, and $p_{22}$ yields
\begin{equation} \label{eqn_33}
	\! \bm{P} \!=\! {\begin{bmatrix}
			p&{-\dfrac{{1.5R}}{d_2} - \dfrac{{{d_1}p}}{{{d_2}}}}\\[4mm]
			{-\dfrac{{1.5R}}{d_2} - \dfrac{{{d_1}p}}{{{d_2}}}}&{\dfrac{{1.5\omega L}}{d_2} + \dfrac{{{d_1}(1.5R + {d_1}p)}}{d_2^2}}
	\end{bmatrix}}.
\end{equation}

\begin{theorem_me}[\textbf{Stability Criterion for WT-GSC Switched System}]
	If there exist a symmetric positive definite matrix $\bm{P} \in {\mathbb{R}^{2 \times 2}}$ given by \eqref{eqn_33}, a constant $\mu>0$, and a $\lambda_1 \in [0,1]$, satisfying the following conditions:
	\begin{equation} \label{eqn_34}
		\!\!\!\begin{cases}
			\! -1.5R{x_{1e}} + 1.5 \omega L{x_{2e}} + v_{{\mathrm{LVRT}}} - v_G = 0 \\[1mm]
			\! \left[ {{\lambda _1}{\bm{A}_1} \!\!+\! ({1 \!\!-\!\! {\lambda _1}}) {\bm{A}_2}} \right] \! {\bm{x}_e} \!\!+\! \left[ {{\lambda _1}{\bm{B}_1}{\bm{u}_1} \!\!+\! ({1 \!\!-\!\! {\lambda _1}}){\bm{B}_2}{\bm{u}_2}} \right] \!=\! \bm{0} \\[1mm]
		\end{cases}
	\end{equation}
	\begin{equation} \label{eqn_35}
		\begin{cases}
			\bm{P} \succ \bm{0} \\[1mm]
			\bm{A}_1^\T \bm{P} + \bm{P}{\bm{A}_1} + \mu \bm{P} \prec \bm{0} \\[1mm]
			\bm{A}_2^\T \bm{P} + \bm{P}{\bm{A}_2} + \mu \bm{P} \prec \bm{0}
		\end{cases}.
	\end{equation}
	Then, the WT-GSC switched system is asymptotically stable about $\bm{x}_e$ during the repeated LVRT switching process.
\end{theorem_me}

The steps for solving this stability problem are explained as follows:

\textit{Step 1: Solve the equation conditions in \eqref{eqn_34}.} This equation group contains three equations and four unknowns (i.e., ${x_{1e}}$, ${x_{2e}}$, $v_{{\mathrm{LVRT}}}$, and $\lambda_1$). The expected stable equilibrium point of the switched system $\left( {{x_{1e}},{x_{2e}}} \right)$ is the quantity to be determined. According to actual control requirements, this problem can be solved by predetermining $v_{{\mathrm{LVRT}}}$ or $\lambda_1$. Generally, $v_{{\mathrm{LVRT}}}$ is given.

\textit{Step 2: Solve the inequality conditions in \eqref{eqn_35}.} This inequality group can be expanded to six linear inequalities about $p$ according to the matrix positive/negative definite criteria. Substituting a specific constant $\mu > 0$ into \eqref{eqn_35}, if this problem is solvable, then the WT-GSC switched system is stable.

For the purpose of fully reflecting the stability differences of the switched system under varying parameter groups, we define the following optimization problem:
\begin{equation} \label{eqn_36}
	\begin{aligned}
		&\max \mu \\
		&s.t. \begin{cases}
			\bm{P} \succ \bm{0} \\[1mm]
			\bm{A}_1^\T \bm{P} + \bm{P}{\bm{A}_1} + \mu \bm{P} \prec \bm{0} \\[1mm]
			\bm{A}_2^\T \bm{P} + \bm{P}{\bm{A}_2} + \mu \bm{P} \prec \bm{0}
		\end{cases}
	\end{aligned}
\end{equation}
where $\mu$ serves as the switched system stability index under a specific parameter group. If $\mu>0$, the system is stable, with larger $\mu$ indicating faster convergence and better stability.


So far, the stability criterion and stability index calculation methods for the wind turbine's repeated LVRT process have been obtained based on the WT-GSC switched system model and its common Lyapunov function.

\section{Sensitivity Analysis and Parameter Optimization}
From Section IV, especially in \eqref{eqn_34} and \eqref{eqn_35}, it is evident that the WT-GSC switched system stability is influenced by many factors, including the wind turbine's external connection impedance and internal control dynamics, which are represented by the line impedance, reactive current proportional coefficient, current inner-loop PI parameters, etc. The analytic expression for the stability index concerning these influencing parameters is hard to obtain. To solve this high-dimensional nonlinear parameter optimization problem for system stability improvement, which is to $\max (\max \mu)$ by adjusting the aforementioned parameters, we employ the Sobol' global sensitivity analysis method to identify dominant parameters, followed by searching for the optimal parameter group via the PSO algorithm in this section.

\subsection{Parameter Sensitivity Analysis Based on the Sobol' Method}
The Sobol' method is a quantitative global sensitivity analysis technique based on variance decomposition, capable of handling nonlinear and non-additive systems \cite{Ref40_SaltelliHowAvoidPerfunctory2010}. The basic principle of the Sobol' method is introduced in this subsection.

The stability index calculation can be viewed as a black box described by a function $Y=g\left(\bm{X}\right)$, where the input vector $\bm{X}=[X_1,X_2,\cdots,X_k]$ considers $k$ of multiple influencing parameters, and the output scalar $Y$ represents the stability index $\mu$. All the input variables $X_i$ are independent and uniformly distributed in a $k$-dimensional unit hypercube $\bm{\Omega}^k = \{\bm{X} | 0 \le X_i \le 1, \ i=1, \cdots, k \}$. Since there exist limited outputs for any points inside $\bm{\Omega}^k$, the function $Y=g\left(\bm{X}\right)$ is continuous and integrable.

The high-dimensional model representation (HDMR) of $g\left(\bm{X}\right)$ with $2^k$ decomposition terms can be expressed as
\begin{equation} \label{eqn_37}
	\begin{aligned}
		g\left(\bm{X}\right) 
		&= g_0 +\sum_{i=1}^k g_i\left(X_i\right) + \hspace{-3.5mm} \sum_{1 \le i<j \le k}^k \hspace{-3.5mm} g_{i,j}\left(X_i, X_j\right) \\
		& \hspace{8mm} + \cdots+g_{1,2, \cdots, k}\left(X_1, X_2, \cdots, X_k\right)
	\end{aligned}
\end{equation}
where $g_0$ is a constant, and other terms are functions only of the input variables in their own indices. Besides, $g\left(\bm{X}\right)$ and all the decomposition terms are integrable \cite{Ref41_SobolGlobalSensitivityIndices2001}.

Since the integrable function $Y=g\left(\bm{X}\right)$ is also square-integrable over $\bm{\Omega}^k$, \eqref{eqn_37} can be squared and integrated to obtain the decomposed variance expression of $Y$ as
%
\begin{equation} \label{eqn_42}
	\begin{aligned}
		{V\!ar}\left(Y\right) = \sum_{i=1}^k {V\!ar}_i + \hspace{-3.5mm} \sum_{1 \le i<j \le k}^k \hspace{-3.5mm} {V\!ar}_{i,j} + \cdots+{V\!ar}_{1,2, \cdots, k}
	\end{aligned}
\end{equation}
where each term represents the partial variance contribution from the corresponding input variable or the interaction of multiple variables.


Generally, the main effect index $S_i$ and the total effect index $S_{Ti}$ are selected to quantify a model's sensitivity with respect to the input variable $X_i$, which can be calculated by
\begin{subequations} \label{eqn_44}
	\begin{align}
		S_i &= \dfrac{{V\!ar}_i}{{V\!ar}} = \dfrac{{V\!ar}_{X_i}\left(E_{\bm{X}_{\sim i}}\left(Y|X_i\right)\right)}{{V\!ar}} \\
		S_{Ti} &= \dfrac{{V\!ar}_i}{{V\!ar}} + \hspace{-3mm} \sum\limits_{1 \le i<j \le k}^k \hspace{-3mm} \dfrac{{V\!ar}_{i,j}}{{V\!ar}} + \cdots + \dfrac{{V\!ar}_{1,2,\cdots,k}}{{V\!ar}} \nonumber \\
		&= \dfrac{E_{\bm{X}_{\sim i}} \left({V\!ar}_{X_i}\left(Y|\bm{X}_{\sim i}\right)\right)}{{V\!ar}} 
	\end{align}
\end{subequations}
where $S_i$ reflects the effect of varying $X_i$ alone on the output variance, $S_{Ti}$ measures the contribution of $X_i$ plus all higher-order effects with $X_i$ to the output variance, ${V\!ar}_{X_i}\left(E_{\bm{X}_{\sim i}} \left(Y|X_i\right)\right)$ represents the expected reduction in variance if $X_i$ could be fixed, and $E_{\bm{X}_{\sim i}} \left({V\!ar}_{X_i}\left(Y|\bm{X}_{\sim i}\right)\right)$ represents the expected value of the left variance if all variables but $X_i$ could be fixed \cite{Ref43_SaltelliGlobalSensitivityAnalysis2008}.

In practice, the Monte Carlo estimator shown in \eqref{eqn_45} is widely used to simultaneously compute indices in \eqref{eqn_44}, which is derived in \cite{Ref45_SaltelliVarianceBasedSensitivity2010}.
\begin{subequations} \label{eqn_45}
	\begin{align}
		{V\!ar}_{X_i}\left(E_{\bm{X}_{\sim i}}\right) & \! \approx \dfrac{1}{M} \sum\limits_{j=1}^M g\left(\bm{B}\right)_j \left( g\left(\bm{A}_{\bm{B}}^i\right)_j - g\left(\bm{A}\right)_j \right) \\
		E_{\bm{X}_{\sim i}} \left({V\!ar}_{X_i}\right) & \! \approx \dfrac{1}{2M} \sum\limits_{j=1}^M \left( g\left(\bm{A}\right)_j - g\left(\bm{A}_{\bm{B}}^i\right)_j \right)^2
	\end{align}
\end{subequations}
where $M$ is the number of sampling points, $\bm{A}$ and $\bm{B}$ are sampling matrices.

\subsection{Parameter Optimization Based on the PSO Algorithm}
The PSO algorithm is a type of heuristic algorithms characterized by stochastic optimization, memory retention, parallel computing, and global searching capabilities \cite{Ref46_KennedyParticleSwarmOptimization1995, Ref47_WuEnergyStorageDevice2014}. Its advantages of fast convergence, few parameters, and simple steps render it suitable for the high-dimensional nonlinear optimization problem investigated in this section.

In this parameter optimization problem, the particle position represents the parameter group corresponding to the fitness value $\mu$, and the particle velocity represents parameter updates causing the change of $\mu$. After stochastic initialization, all particles iteratively update their velocities $\bm{v}$ and positions $\bm{z}$ by tracking personal and global best positions, expressed as
\begin{equation} \label{eqn_46}
	\begin{cases}
		\bm{v}_{i}^{m+1} = w^m\bm{v}_{i}^m \!+\! c_1 \bm{r}_{1,i}^m \left(\bm{z}_{pb,i}^m \!-\! \bm{z}_{i}^m\right) \!+\! c_2 \bm{r}_{2,i}^m \left(\bm{z}_{gb,i}^m \!-\! \bm{z}_{i}^m\right) \\
		\bm{z}_{i}^{m+1} = \bm{z}_{i}^{m} + \bm{v}_{i}^{m+1}
	\end{cases}
\end{equation}
where the superscript $m$ denotes the $m$th iteration, the subscript $i$ denotes the $i$th particle, $w$ is the inertia weight, $c_1$ and $c_2$ are acceleration coefficients, $\bm{r}_1$ and $\bm{r}_2$ are vectors comprising random numbers in the interval $\left(0,1\right)$, and $\bm{z}_{pb}$ and $\bm{z}_{gb}$ represent position vectors corresponding to the particles' personal and global best values of $\mu$, respectively.

The PSO algorithm terminates when $m$ reaches the maximum number of iterations or the fitness value $\mu$ converges, obtaining the global optimal solution for the parameter group and corresponding fitness value.

\section{Simulation Verification}
\begin{figure}[!htbp]
	\centering
	\includegraphics[width=7cm]{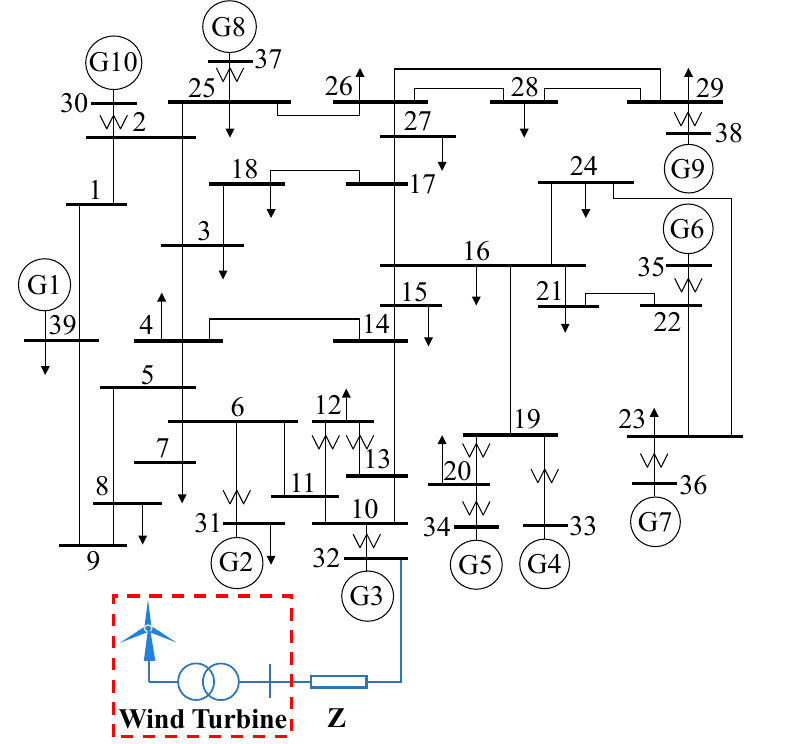}
	\caption{Topology of the modified IEEE 39-bus test system.}
	\label{fig_5}
\end{figure}

In this section, the effectiveness of the proposed WT-GSC switched system model, stability analysis approaches, sensitivity analysis methods, and parameter optimization algorithms are demonstrated through a case study on the electromagnetic transient simulation platform — CloudPSS \cite{Ref48_SongCloudPSSHighperformancePower2020, Ref49_CloudPSS}. As shown in Fig. \ref{fig_5}, the test system is a modified IEEE New England 39-bus system with a 1.5MW wind turbine connected to bus 32 through a 0.69/20kV transformer and an equivalent impedance. The connection impedance and control parameters of the wind turbine are the same as those in Table \ref{table_1}, and the base frequency and simulation step are set to 50Hz and 50ms, respectively.

\begin{figure}[!htbp]
	\centering
	\includegraphics[width=7cm]{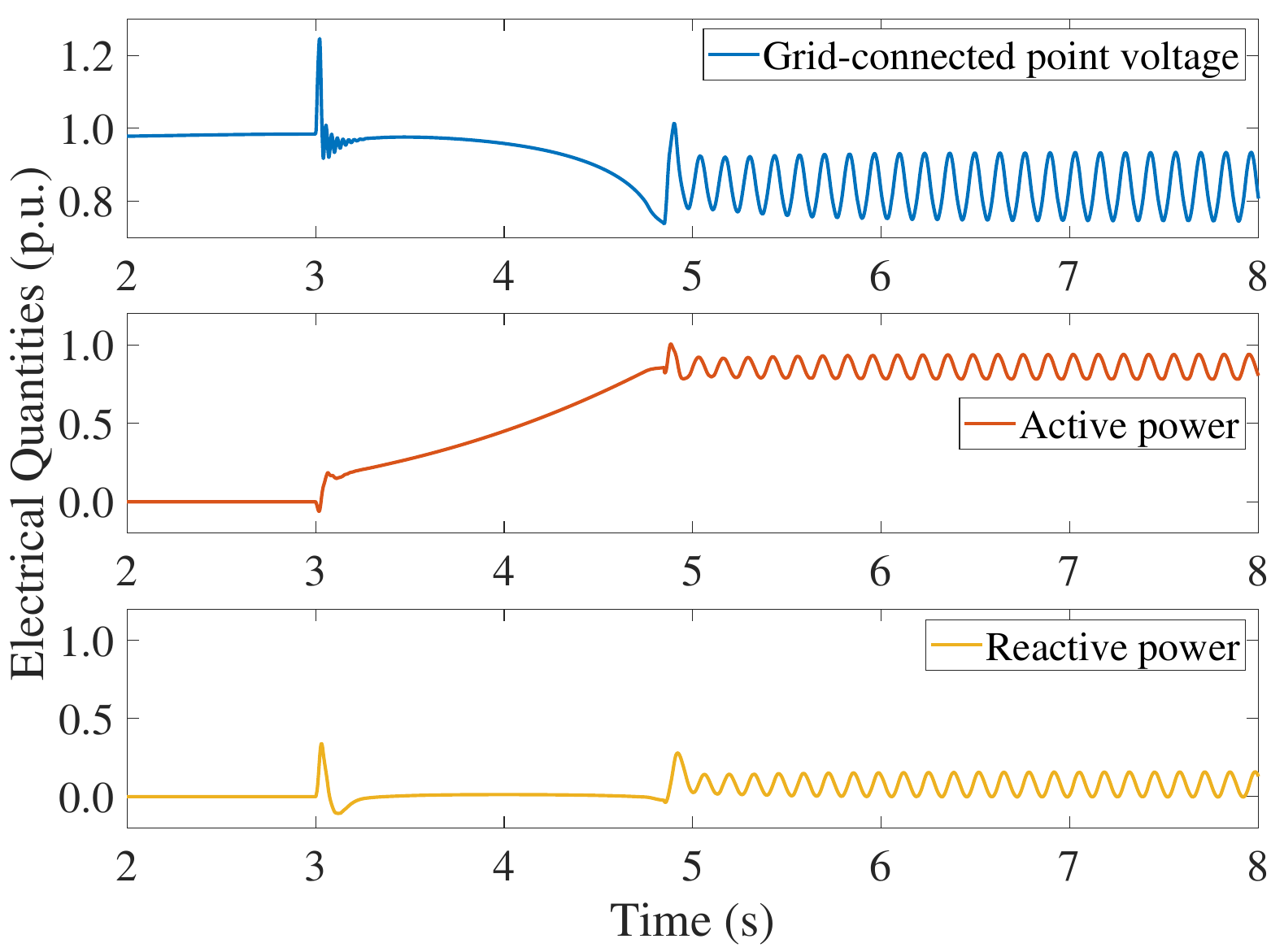}
	\caption{Simulation reproduction of the voltage oscillation phenomenon induced by repeated LVRT of the wind turbine.}
	\label{fig_6}
\end{figure}

The SCR of the test system measuring at the wind turbine's grid-connected point is approximately 1.84, indicating this is a weak system. As shown in Fig. \ref{fig_6}, when the active power output of the wind turbine increases, the grid-connected point voltage decreases until becomes lower than the LVRT threshold (here we take $v_{{\mathrm{LVRT}}}$ as 0.80p.u. \cite{Ref15_GBT_wind_farm_specification_2021}), thereby inducing repeated LVRT voltage oscillations according to the mechanism proposed in Section III.

\subsection{Verification of Modeling Assumptions}
\begin{figure}[!htbp]
	\centering
	\vspace{-3mm}
	\subfloat[]{
		\includegraphics[width=4.2cm]{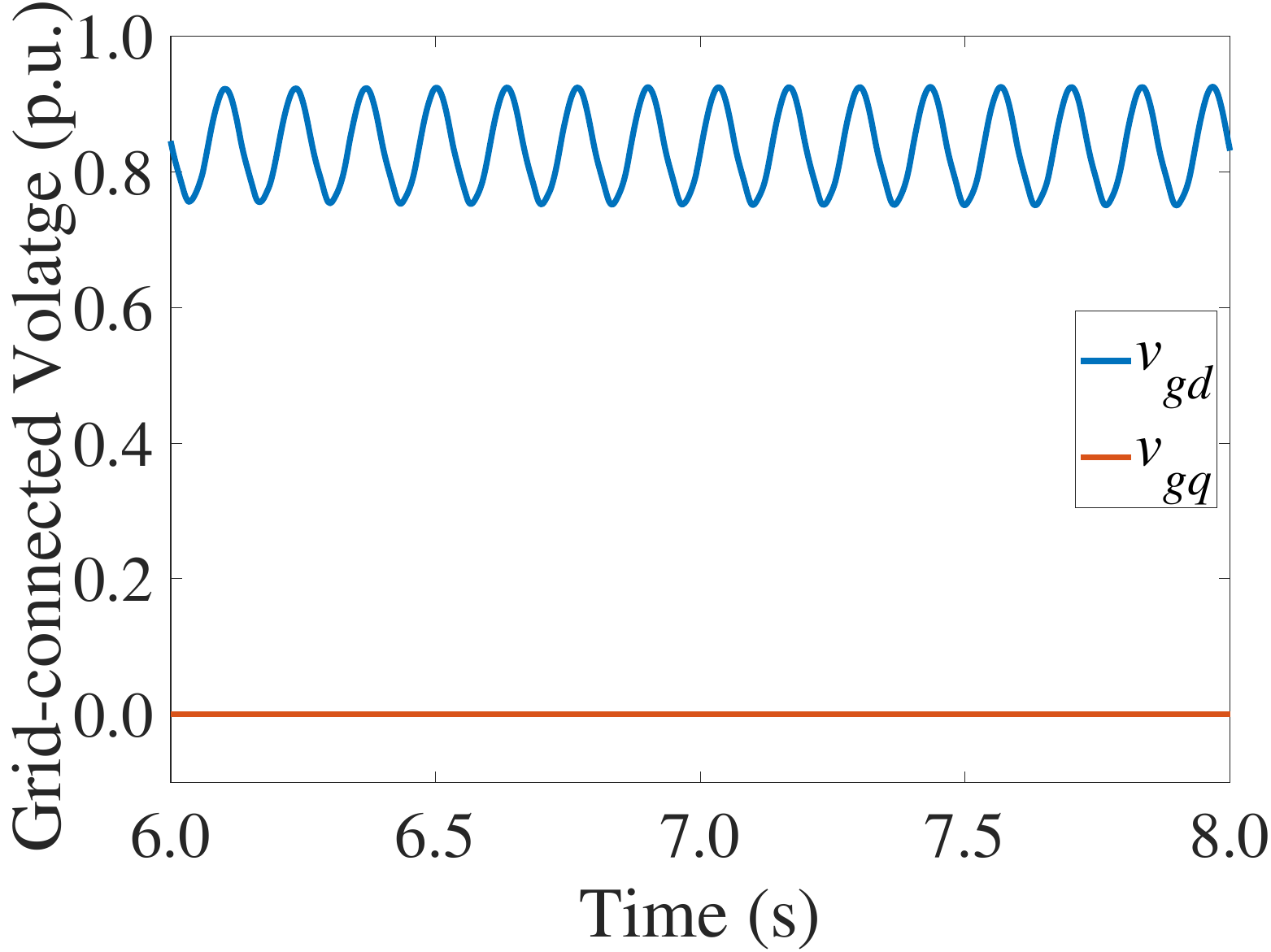}
		\label{fig_7a}
	}
	\subfloat[]{
		\includegraphics[width=4.2cm]{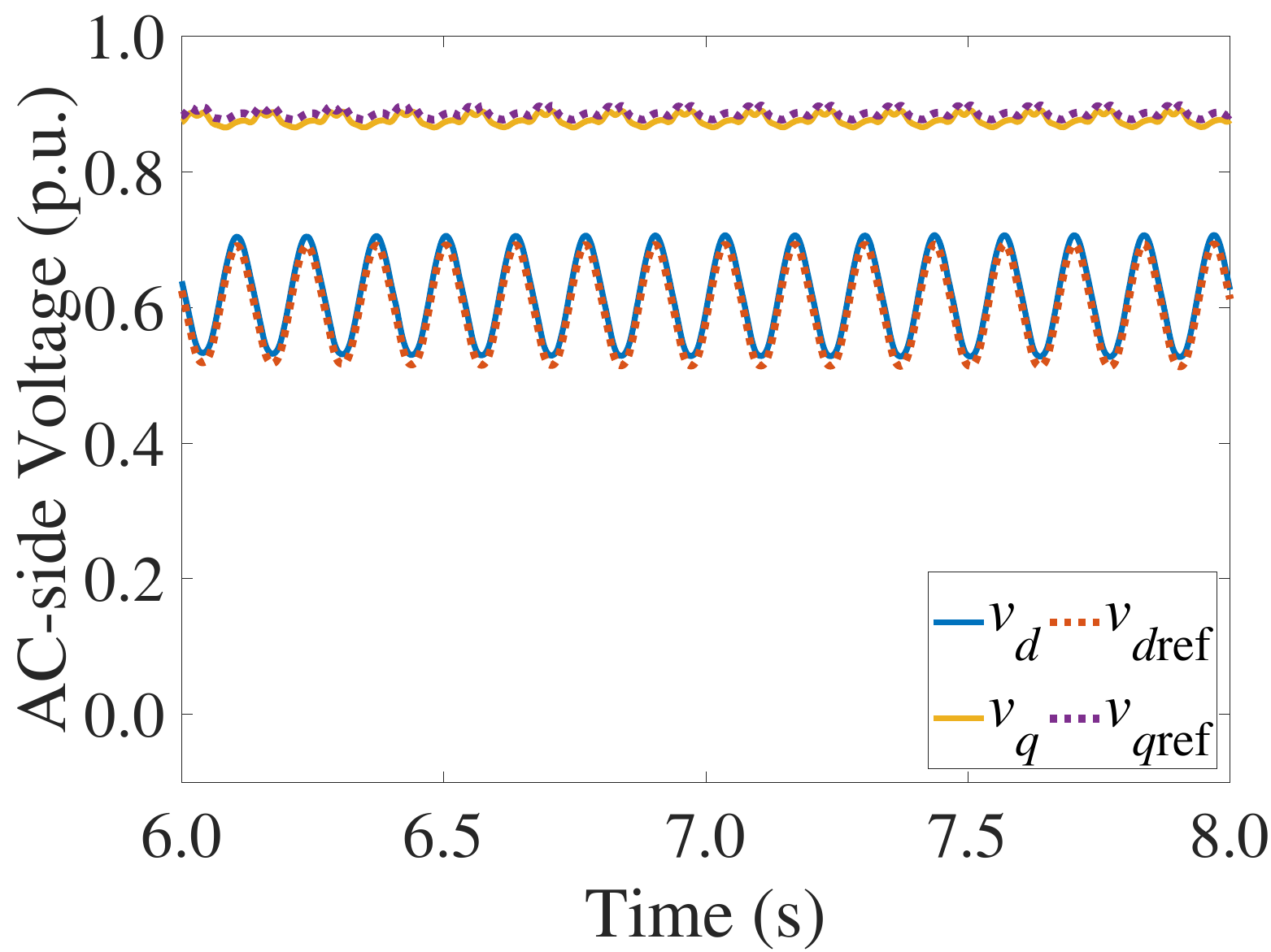}
		\label{fig_7b}
	} \\[-3mm]
	\subfloat[]{
		\includegraphics[width=4.2cm]{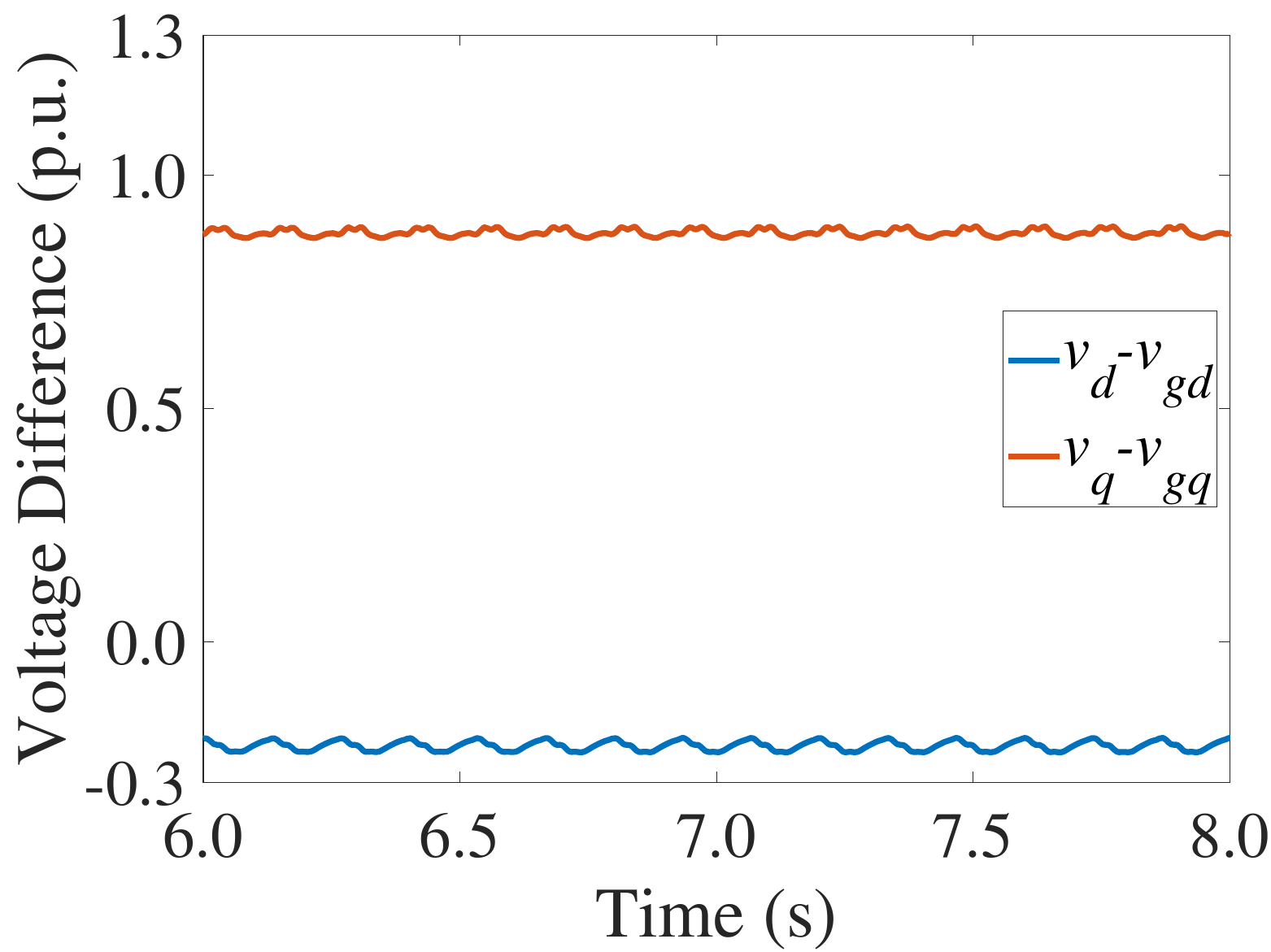}
		\label{fig_7c}
	}
	\subfloat[]{
		\includegraphics[width=4.2cm]{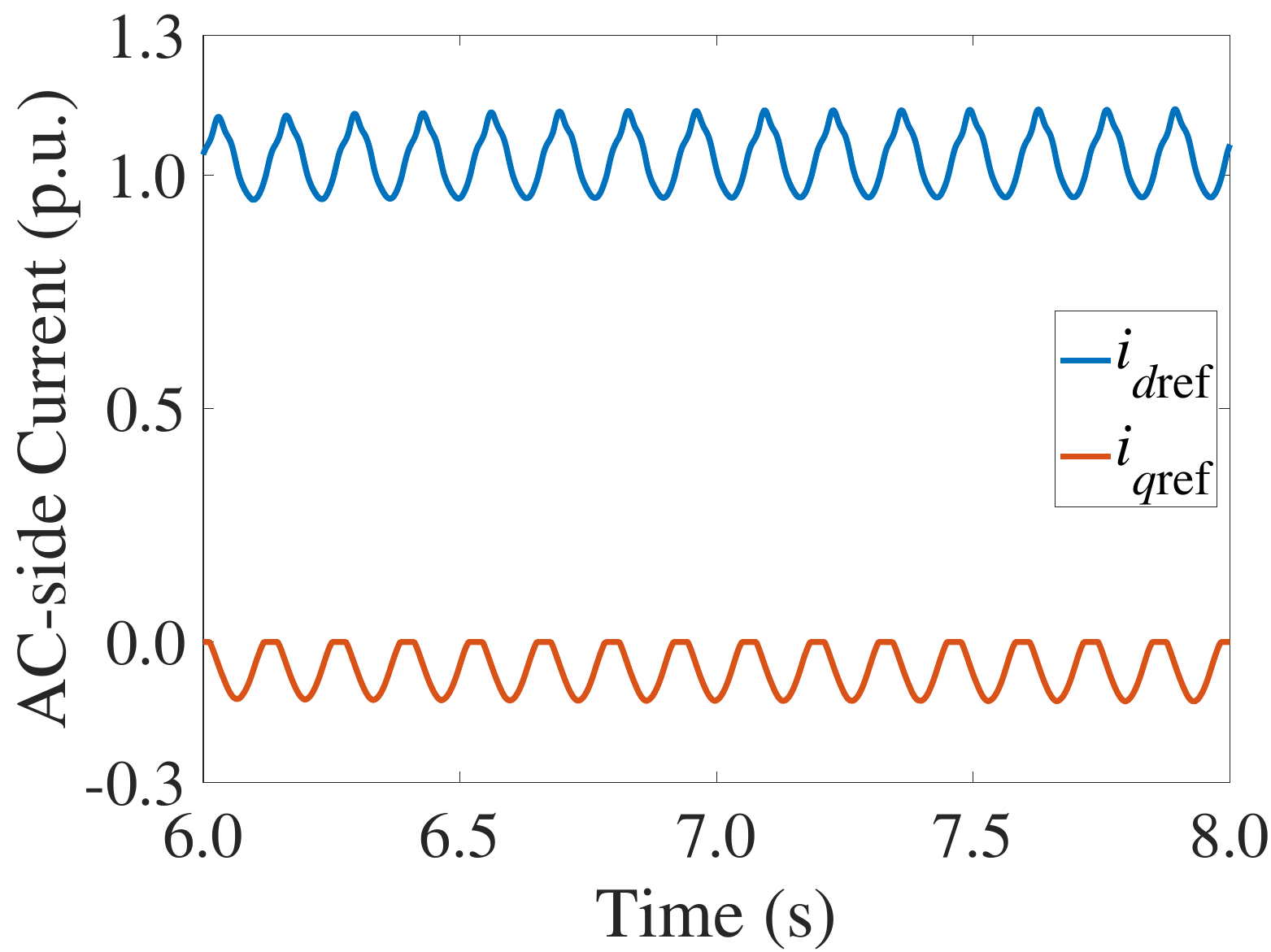}
		\label{fig_7d}
	}
	\caption{Simulation verification of modeling assumptions. (a) The actual value of grid-connected point voltage. (b) The actual and reference values of AC-side voltage, where the solid lines represent actual values and the dashed lines represent reference values. (c) The difference value between AC-side and grid-connected point voltage. (d) The reference value of AC-side current.}
	\label{fig_7}
\end{figure}

In the modeling procedure, the WT-GSC switched system is reasonably simplified on the basis of a series of assumptions, which are verified through time-domain simulations, as illustrated in Fig. \ref{fig_7}.

{ 
	\newtheoremstyle{assumption_me}{3pt}{3pt}{\normalshape}{0cm}{\bfseries}{:}{0.2em}{}
	\theoremstyle{assumption_me}
	\newtheorem{assumption_me}{\indent Assumption}
}
\begin{assumption_me}
	The WT-GSC generally adopts grid-voltage-oriented vector control. Fig. \ref{fig_7a} confirms that the \textit{d}- and \textit{q}-axes components of the grid-connected point voltage satisfy ${v_{gd}} = {v_g}$ and ${v_{gq}} = 0$, respectively.
\end{assumption_me}

\begin{assumption_me}
	The voltage PWM control response of the WT-GSC controlled inverter bridge is fast enough to make the AC-side voltage quickly track its reference value, i.e., ${v_{d{\mathrm{ref}}}} \approx {v_d}$ and ${v_{q{\mathrm{ref}}}} \approx {v_q}$. In Fig. \ref{fig_7b}, it can be seen that the actual and reference values are basically the same, with a maximum difference of about 0.014p.u.
\end{assumption_me}

\begin{assumption_me}
	The difference between the AC-side and grid-connected point voltage of the WT-GSC generally fluctuates little and can be approximated as a constant. As shown in Fig. \ref{fig_7c}, the maximum fluctuation is about 0.032p.u., which can be neglected.
\end{assumption_me}

\begin{assumption_me}
	The WT-GSC capacity is large enough so that the AC-side current reference values for both the normal operation and LVRT subsystems are within the current limit. Fig. \ref{fig_7d} verifies that the AC-side current reference value satisfies $\max \sqrt {i_{d{\mathrm{ref}}}^2 + i_{q{\mathrm{ref}}}^2} \approx 1.15\mathrm{p.u.} < {I_{\max }}$.
\end{assumption_me}

\subsection{Verification of Voltage Oscillation Evolution Dynamics}
\begin{figure}[!htbp]
	\centering
	\vspace{-3mm}
	\subfloat[]{
		\includegraphics[width=4.2cm]{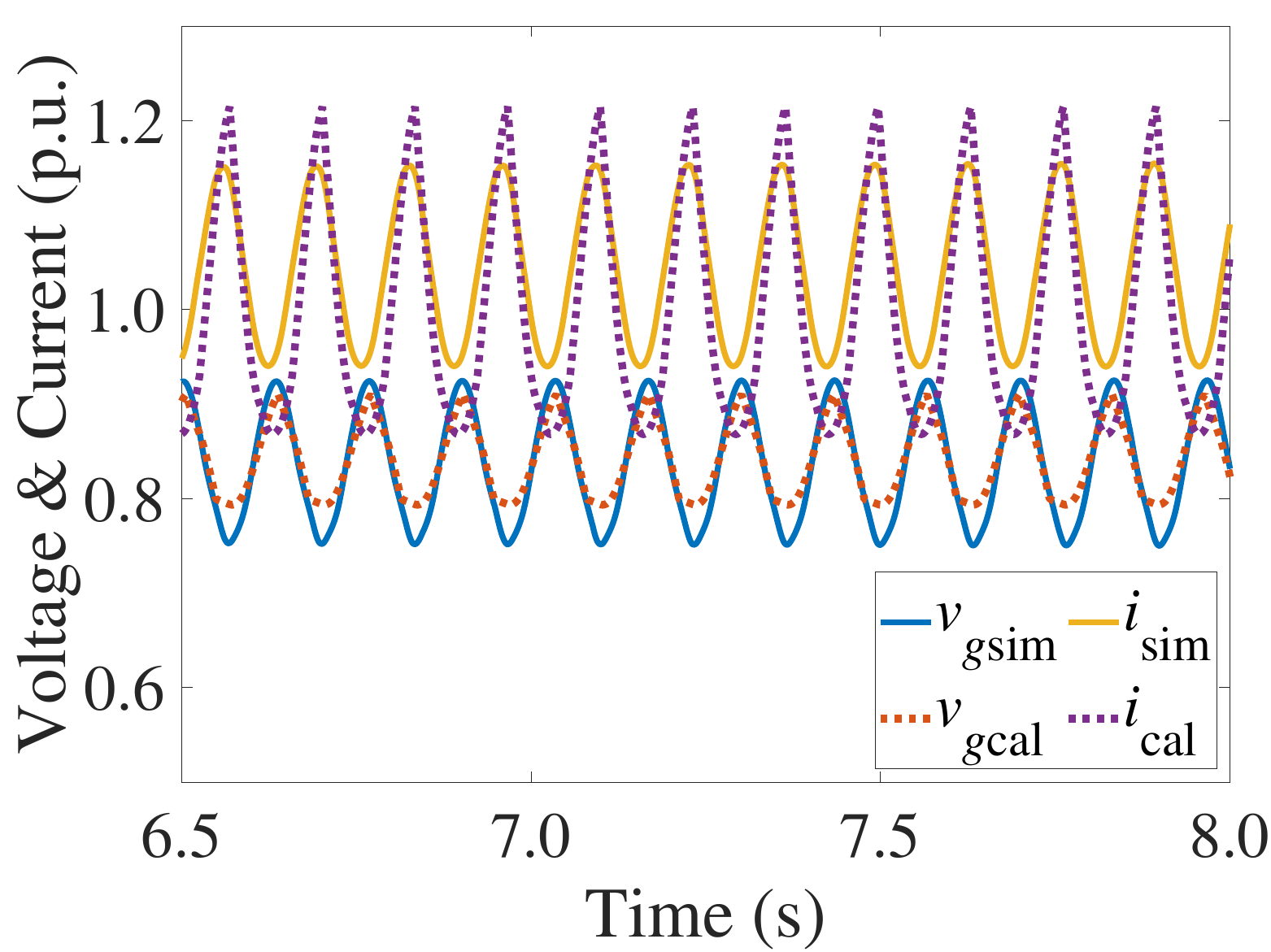}
		\label{fig_8a}
	}
	\subfloat[]{
		\includegraphics[width=4.2cm]{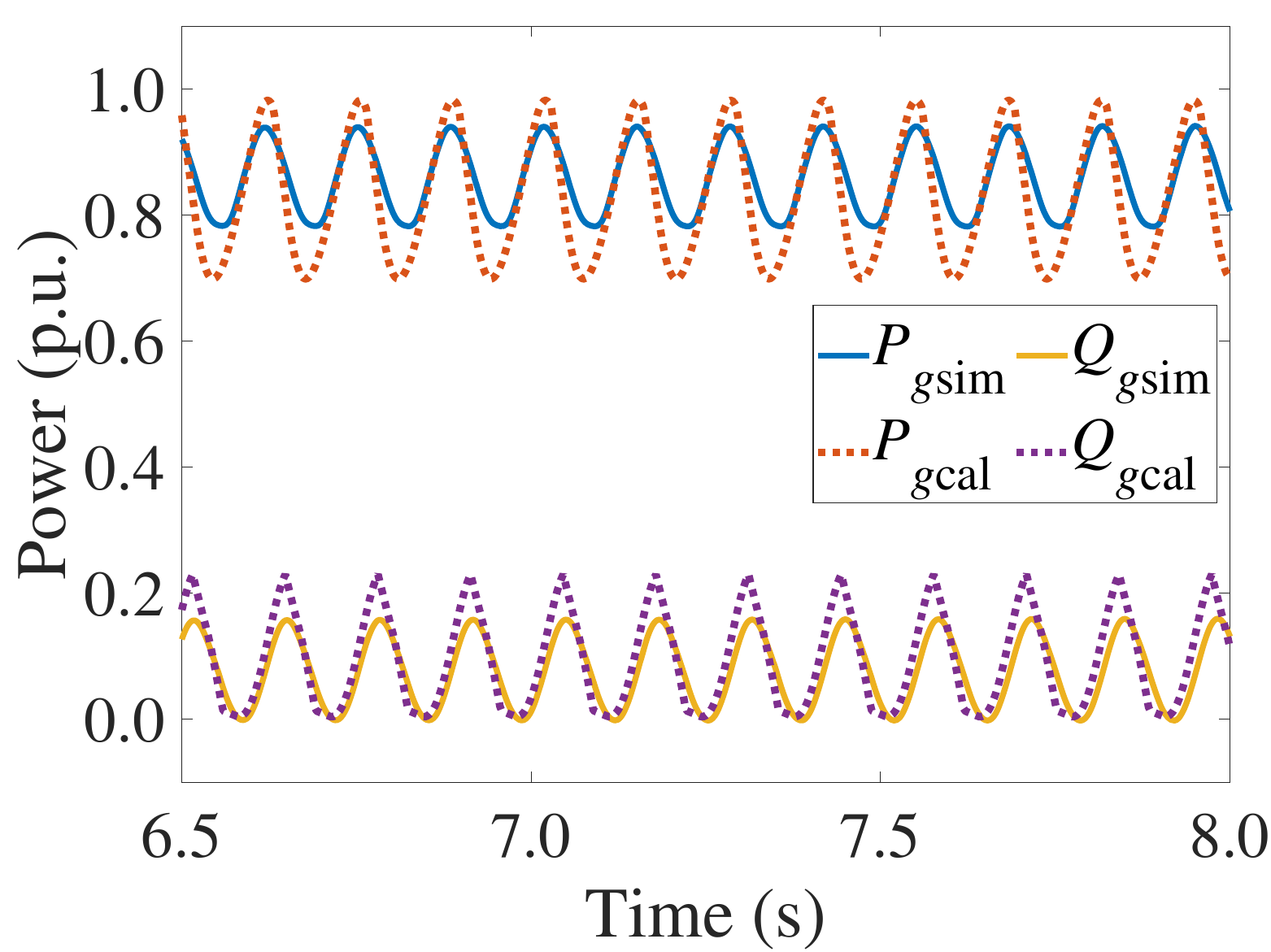}
		\label{fig_8b}
	}
	\caption{Comparison between the simulation and calculation values for electrical quantities during the repeated LVRT dynamic process of the wind turbine. (a) Voltage and current curves. (b) Active and reactive power curves. The subscript ``sim'' in the legend indicates simulation values, corresponding to solid lines; ``cal'' indicates calculation values, corresponding to dashed lines.}
	\label{fig_8}
\end{figure}

The voltage oscillation phenomenon caused by the wind turbine repeatedly entering and exiting LVRT is reproduced through simulation, verifying the dynamic evolution process of the grid-connected point voltage, current, and power characterized by the WT-GSC switched system model. The comparison between simulation and calculation results is shown in Fig. \ref{fig_8}.

As can be seen from Fig. \ref{fig_8}, the theoretical calculation results of the WT-GSC switched system model and simulation results of the test system for the grid-connected point voltage, current, active power output, and reactive power output during the repeated LVRT process exhibit strong concordance in both oscillation frequency and transient behavior. Computed over an oscillation period (e.g., 6.500$\sim$6.634s), the root-mean-square errors are approximately 0.022, 0.068, 0.080, and 0.045, respectively, all below 10\%. Furthermore, the prediction errors of the grid-connected point voltage and reactive power are less than 5\%. These results validate the correctness and effectiveness of the WT-GSC switched system model.

\subsection{Effectiveness of Stability Criterion and Stability Index}
In order to verify the effectiveness of the proposed stability criterion and stability index for the repeated LVRT process of the wind turbine, calculations and simulations on multiple parameter groups of stable or unstable conditions should be conducted. The quasi-Monte Carlo method is employed to generate low-discrepancy sequences (e.g., the Sobol' sequence) \cite{Ref45_SaltelliVarianceBasedSensitivity2010} for random sampling of parameters within the ranges as listed in Table \ref{table_2}.

\begin{table}[!htbp]
	\caption{Parameter Ranges for Random Sampling}
	\setstretch{0.9}
	\label{table_2}
	\centering
	\resizebox{0.9\columnwidth}{!}{
	\begin{tabular}{cccc}
		\toprule
		\toprule
		Parameter & Range (p.u.) & Parameter & Range (p.u.) \\
		\midrule
		$K_{pd}$, $K_{pq}$ & 0.10 $\sim$ 0.20 & $L_g$ & 3.25e-4 $\sim$ 5.41e-4 \\[1mm]
		$K_{id}$, $K_{iq}$ & 1.00 $\sim$ 5.00 & $R$   & 2.17e-4 $\sim$ 4.50e-3 \\[1mm]
		$K_1$              & 1.50 $\sim$ 3.00 & $L$   & 8.66e-4 $\sim$ 1.10e-3 \\
		\bottomrule
		\bottomrule
	\end{tabular}
	}
\end{table}

For all the sampling points, the stability analysis results obtained from the analytical model calculation and test system simulation are counted and compared in Table \ref{table_3}, which show that the precision and recall \cite{Ref50_ZhouZhihuaMachineLearning_2016} of stability determination given by the proposed stability criterion can exceed 85.0\% and 80.0\%, respectively. The errors may come from the modeling simplification and conservativeness of the criterion.

\begin{table}[!htbp]
	\setstretch{0.9}
	\caption{Accuracy Verification of the Stability Criterion}
	\label{table_3}
	\centering
	\resizebox{\columnwidth}{!}{
	\begin{threeparttable}
		\begin{tabular}{ccccc}
			\toprule
			\toprule
			\makecell[c]{Sampling \\ Number} & \makecell[c]{Valid Case \\ Number\tnote{1}} & \makecell[c]{Stable Case Number \\ (Both) (Cal.) (Sim.)}& Precision\tnote{2} & Recall\tnote{3} \\
			\midrule
			32  & 21 & 12 \quad \ 14 \quad \ 15 & 85.7\% & 80.0\% \\[1mm]
			64  & 38 & 24 \quad \ 27 \quad \ 29 & 88.9\% & 82.8\% \\[1mm]
			128 & 79 & 46 \quad \ 54 \quad \ 57 & 85.2\% & 80.7\% \\
			\bottomrule
			\bottomrule
		\end{tabular}
	 	\begin{tablenotes}
			\footnotesize
			\item[1] Excluding the voltage collapse cases gives the valid cases.
			\item[2] The precision is calculated by diving the number of stable cases in both calculation and simulation by the number of stable cases in calculation.			
			\item[3] The recall is calculated by diving the number of stable cases in both calculation and simulation by the number of stable cases in simulation.
		\end{tablenotes}
	\end{threeparttable}
	}
\end{table}

\begin{figure}[!htbp]
	\centering
	\includegraphics[width=6.5cm]{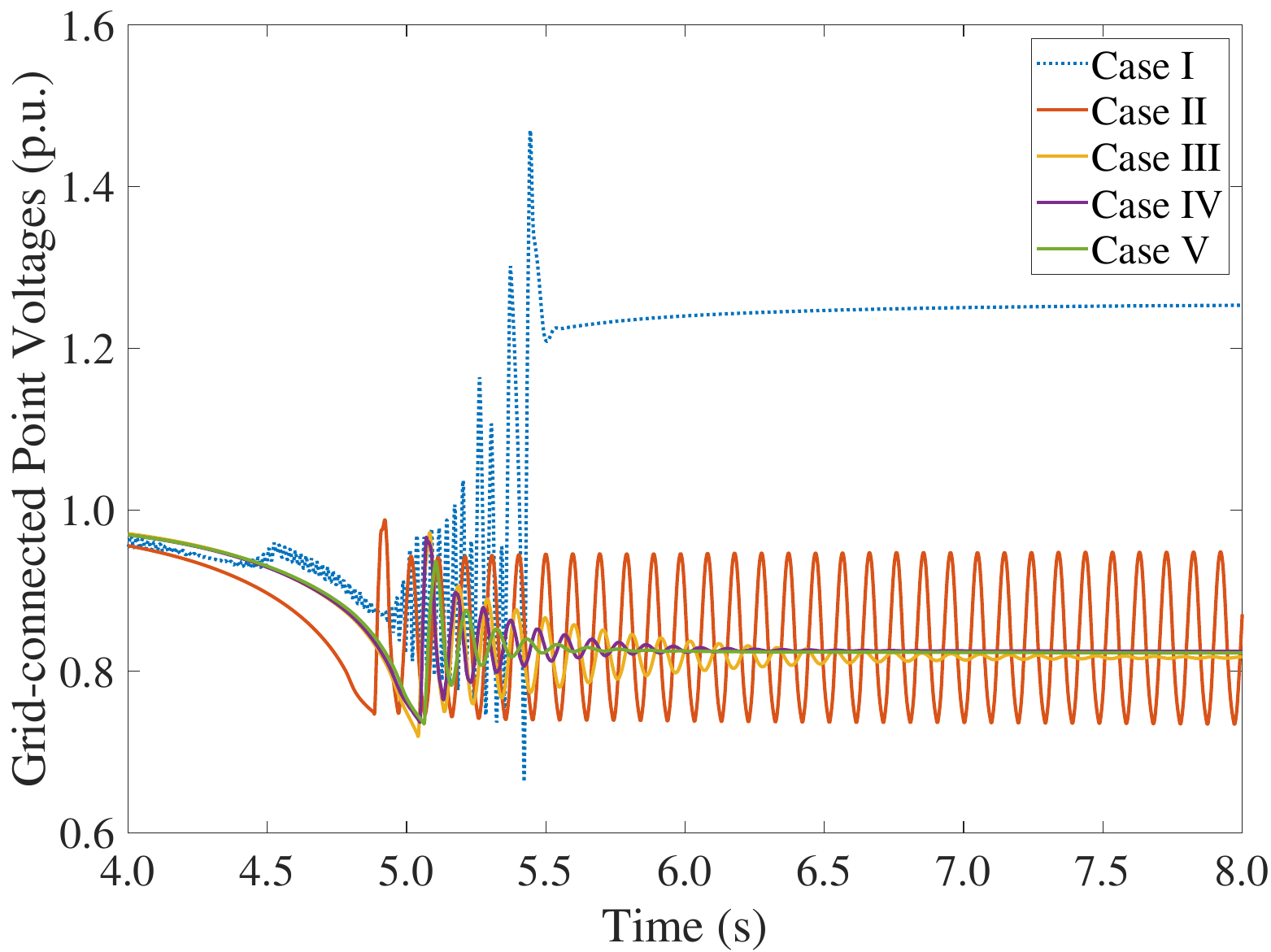}
	\caption{Grid-connected point voltages under different parameter groups. Case I: voltage collapse. Case II: $\mu = -20.51$, unstable. Case III: $\mu = 0.40$, slowly stable. Case IV: $\mu = 6.05$, stable. Case V: $\mu = 22.36$, fast stable.}
	\label{fig_9}
	\vspace{-3mm}
\end{figure}

To illustrate the effectiveness of the stability index, Fig. \ref{fig_9} depicts the stability analysis results of 5 representative cases selected from 128 sampling points. It can be seen that the stability index can reflect the stability degree when applying different parameter groups. The larger $\mu$ is, the faster the system reaches stability and the better the stability is.

\subsection{Sensitivity Analysis and Parameter Optimization}
\begin{figure}[!htbp]
	\centering
	\subfloat[]{
		\includegraphics[width=4.25cm]{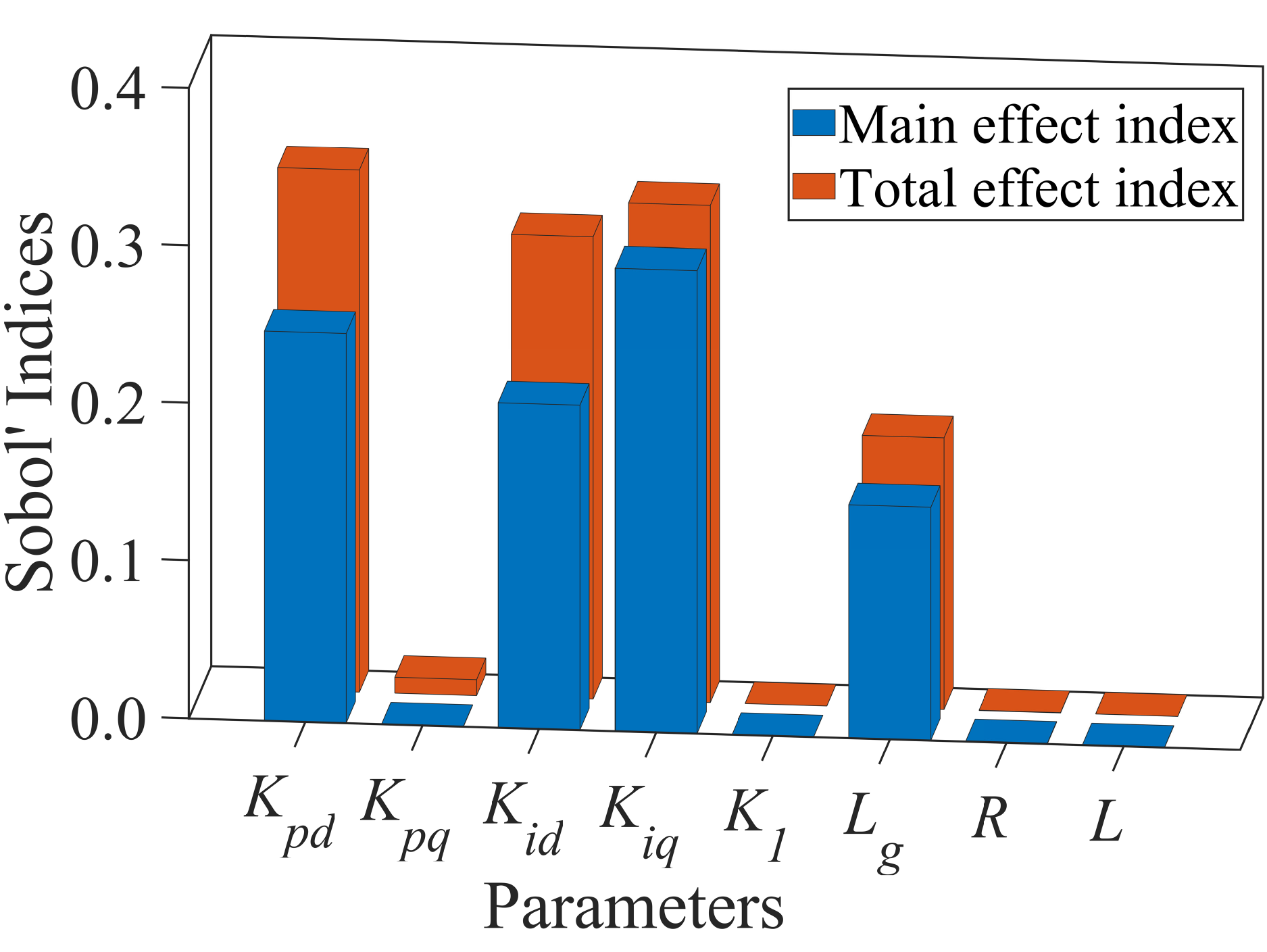}
		\label{fig_10a}
	}
	\subfloat[]{
		\includegraphics[width=4.2cm]{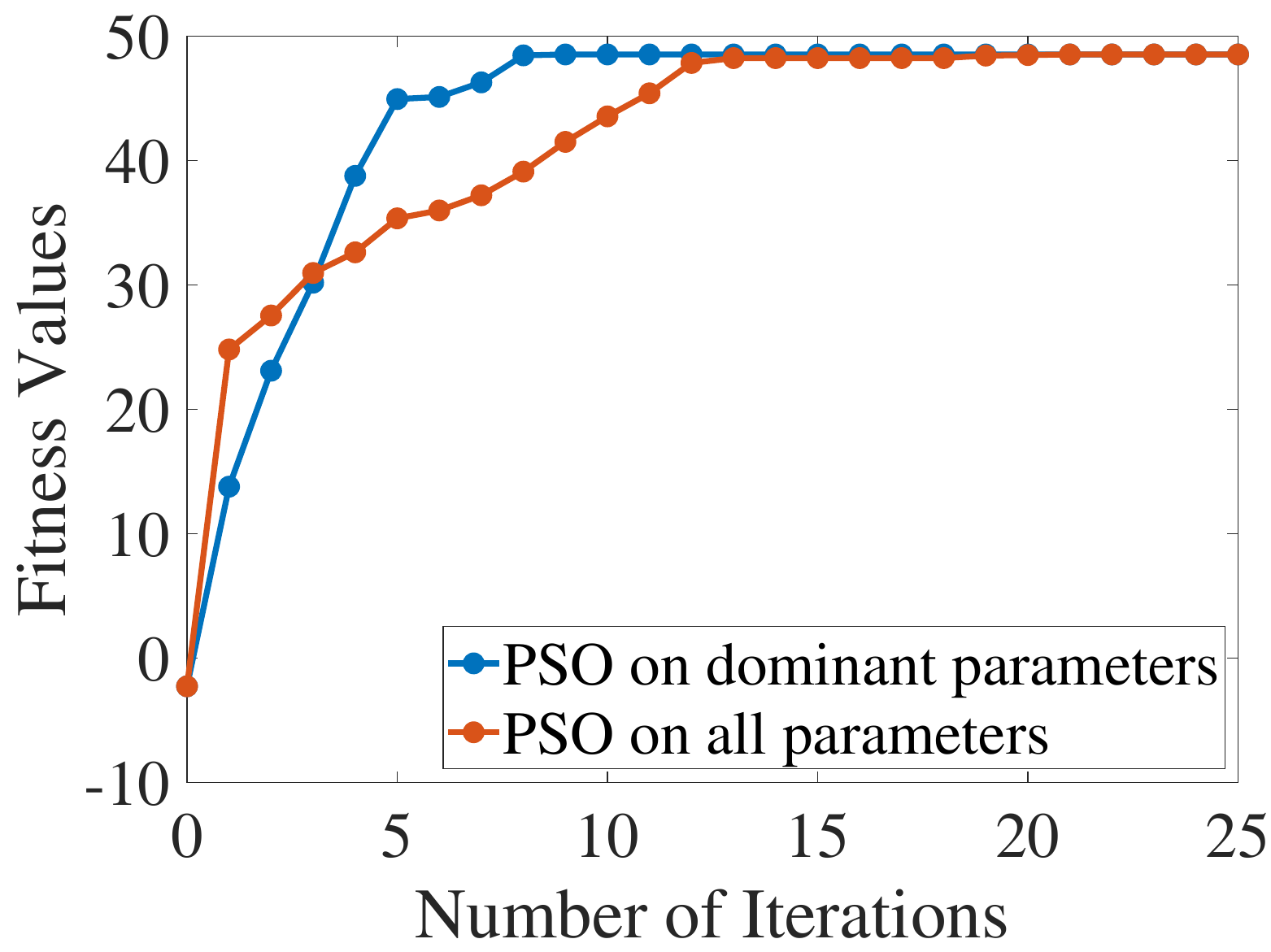}
		\label{fig_10b}
	} \\
	\subfloat[]{
		\includegraphics[width=6.5cm]{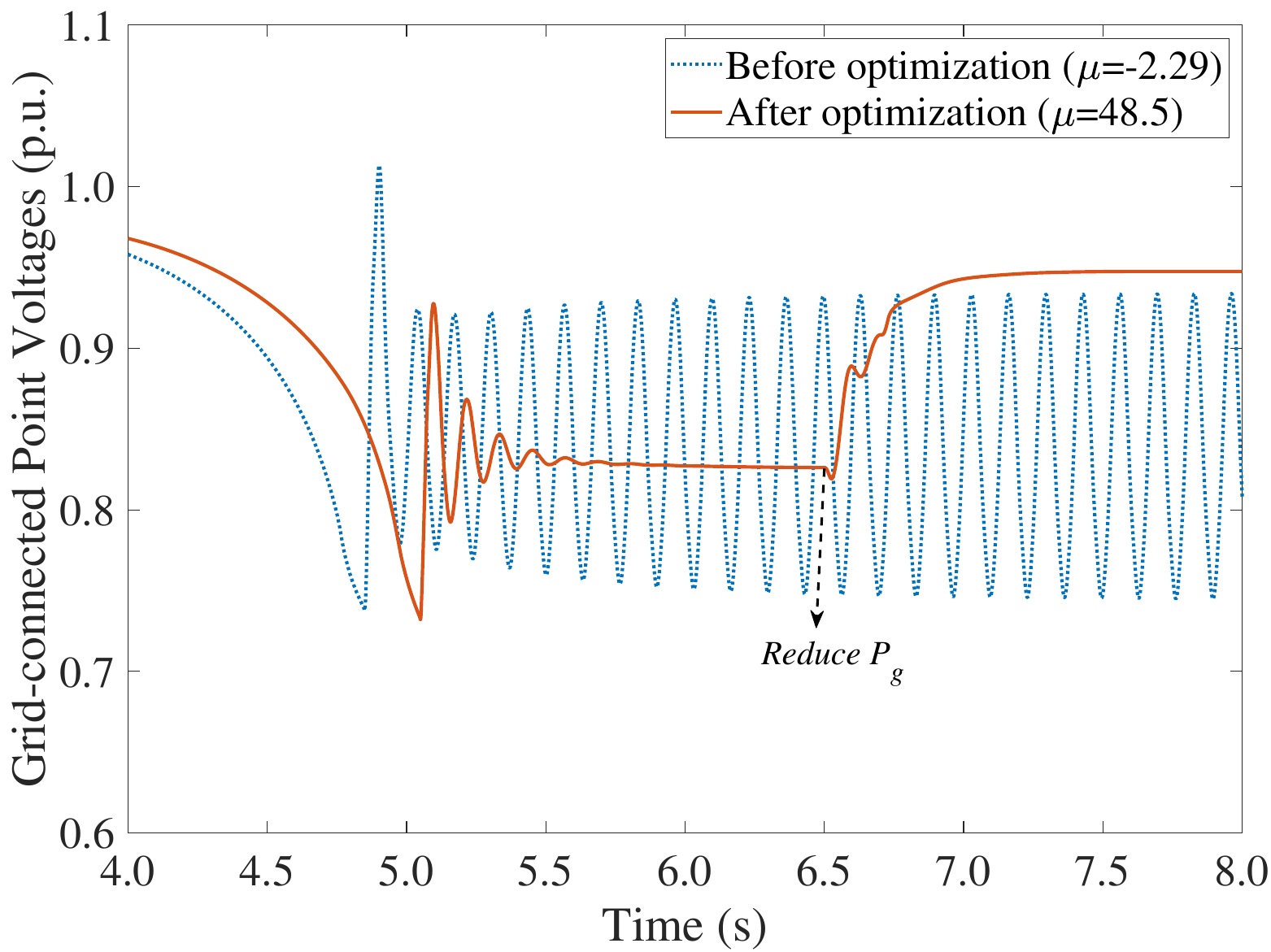}
		\label{fig_10c}
	}
	\caption{Dominant parameter identification and optimization for system stability improvement. (a) Sobol' indices for parameters of the WT-GSC switched system. (b) Fitness value convergence curve based on the PSO. (c) Grid-connected point voltages before and after parameter optimization.}
	\label{fig_10}
	\vspace{-1mm}
\end{figure}

To efficiently improve the stability of the WT-GSC switched system, the Sobol' global sensitivity analysis method is adopted to identify dominant parameters, which can be further optimized via the PSO algorithm, as shown in Fig. \ref{fig_10}.

%

Fig. \ref{fig_10a} reveals that within the parameter ranges in Table \ref{table_2}, $K_{pd}$, $K_{id}$, $K_{iq}$, and $L_g$ have higher Sobol' indices. Thus, they are dominant parameters affecting the system stability. In Fig. \ref{fig_10b}, the optimal fitness value is $\mu =48.5$, corresponding to the parameter group listed in Table \ref{table_4}, which results in the best stability performance of the WT-GSC switched system. In particular, the PSO on dominant parameters takes fewer iterations and less time to converge compared with optimization on all parameters while still achieving satisfactory results. Fig. \ref{fig_10c} shows that after optimization, the voltage fluctuation is smaller, and the system becomes fast stable, validating the effectiveness of parameter optimization. However, due to the weak grid characteristic and large amounts of active power output, the system is stabilized at a low voltage level. Enhancing the grid strength or reducing the active power output would restore the voltage to normal levels.

\begin{table}[!htbp]
	\setstretch{0.9}
	\caption{Parameter Group for Largest Stability Index}
	\label{table_4}
	\centering
	\resizebox{\columnwidth}{!}{
		\begin{tabular}{ccccccccc}
			\toprule
			\toprule
			$K_{pd}$ & $K_{pq}$ & $K_{id}$ & $K_{iq}$ & $K_1$ & $L_g$ & $R$ & $L$ & Unit \\
			\midrule
			0.20 & 0.10 & 5.00 & 5.00 & 1.68 & 3.25e-4 & 1.20e-3 & 8.66e-4 & p.u. \\
			\bottomrule
			\bottomrule
		\end{tabular}
	}
\end{table}

\section{Conclusion}
This paper focuses on a new type of voltage oscillation phenomenon observed in weak grid-connected wind power systems, which is induced by the wind turbine repeatedly entering and exiting LVRT. Switched system theory is introduced for dynamic modeling, mechanism elucidation, and stability analysis of the repeated LVRT process. Specifically, a novel WT-GSC switched system model considering the external connection impedance and internal control dynamics of the wind turbine is constructed, which can effectively characterize the evolution dynamics of various electrical quantities. Moreover, the mechanism underlying this voltage oscillation phenomenon can be explained as the repeated switching of the switched system trajectory between the normal operation and LVRT subsystems under reactive power switching control. Subsequently, the stability analysis and assessment methods are proposed based on the common Lyapunov function, providing a criterion and an index for switched system stability analysis. Furthermore, in order to efficiently enhance system stability, the Sobol' global sensitivity analysis method is adopted to identify the dominant parameters, which are then optimized via the PSO algorithm to find the best parameter group. In the case study of a modified IEEE 39-bus test system on the CloudPSS platform, the effectiveness of the proposed WT-GSC switched system model, the stability analysis and assessment methods, and the sensitivity analysis and parameter optimization approaches are verified.
 
\bibliographystyle{IEEEtran}
\bibliography{References}

\end{document}